\let\label\orilabel
\documentclass[aps,nofootinbib,amsmath,prd,onecolumn,notitlepage,showpacs,superscriptaddress,10pt]{revtex4-1}
\let\orilabel\label

\usepackage[american]{babel}
\usepackage{amsfonts}
\usepackage{amsmath}
\usepackage{amssymb}
\usepackage{epsfig}
\usepackage{latexsym}
\usepackage{dsfont}

\usepackage{fancyhdr}
\usepackage{graphicx}

\usepackage{hyperref}
\usepackage{etex}
\allowdisplaybreaks
\usepackage{float}
\usepackage{slashed}
\usepackage{xcolor}

\usepackage{mathtools}

\usepackage{dcolumn} 
\usepackage{bm}
\usepackage{splitidx}

\usepackage{slashed}
\usepackage{mathrsfs}

\usepackage{fancyhdr}
\makeatletter
\renewcommand{\c@secnumdepth}{0}
\makeatother

\usepackage{hyperref}

\fancypagestyle{plain}{
	\lhead{}
	\fancyhead[R]{\thepage}
	\fancyhead[L]{}
	
	\fancyfoot{}
}

\usepackage{empheq}
\usepackage[most]{tcolorbox}

\newtcbox{\mymath}[1][]{%
	nobeforeafter, math upper, tcbox raise base,
	enhanced, colframe=blue!30!black,
	colback=blue!30, boxrule=1pt,
	#1}

\newcommand{\onabla}{\mathring{\nabla}}
\newcommand{\oGamma}{\mathring{\Gamma}}
\newcommand{\oD}{\mathring{D}}
\newcommand{\oK}{\mathring{K}}
\newcommand{\Q}{\mathbb{Q}}
\newcommand{\barQ}{\overline{\mathbb{Q}}}
\newcommand{\barq}{\overline{q}}

\begin{document}

	\title{Extrinsic geometry and Hamiltonian analysis of symmetric teleparallel gravity}

	\author{Salvatore Capozziello}
	\email{capozziello@na.infn.it}
	\affiliation{Dipartimento di Fisica E. Pancini, Universita di Napoli Federico II Complesso
		Universitario di Monte Sant Angelo, Edificio G, Via Cinthia, I-80126, Napoli, Italy}
	\affiliation{Scuola Superiore Meridionale, Via Mezzocannone 4, I-80134, Napoli, Italy}
	\affiliation{Istituto Nazionale di Fisica Nucleare (INFN), Sez. di Napoli Complesso
		Universitario di Monte Sant Angelo, Edificio G, Via Cinthia, I-80126, Napoli, Italy}
	
	\author{Dario Sauro}
	\email{dario.sauro@uni-jena.de}
	\affiliation{
		Theoretisch-Physikalisches Institut, Friedrich-Schiller-Universität Jena,\\
		Fröbelstieg 1, 07743 Jena, Germany
	}

	\begin{abstract}
		We analyze the properties of foliations in presence of non-metricity, deriving the generalized Gauss-Codazzi relations in full generality. These  results are employed to study the teleparallel framework of non-metric geometry, obtaining  constraints on the extrinsic and intrinsic tensors. In particular, an extrinsic symmetric two-tensor plays the role of the extrinsic curvature in Riemannian geometry, whereas no other geometric object can induce new dynamical degrees of freedom. Furthermore, we analyze the variational principle in  presence of non-metricity, obtaining the boundary terms for the well-posed and well-defined Cauchy problem. Finally, we exploit the previous results to construct the Hamiltonian of the symmetric teleparallel equivalent of General Relativity, providing a proof that this theory shares the same number of degrees of freedom with its Riemannian counterpart.
	\end{abstract}

	\maketitle

	\newcommand{\hD}{\hat{D}}
	\newcommand{\hF}{\hat{F}}
	\newcommand{\hnabla}{\hat{\nabla}}
	\newcommand{\hG}{\hat{G}}
	\newcommand{\hN}{\hat{N}}
	\newcommand{\ha}{\hat{a}}
	\newcommand{\hOmega}{\hat{\Omega}}
	\newcommand{\hM}{\hat{M}}
	\newcommand{\hK}{\hat{K}}
	\newcommand{\hY}{\hat{Y}}
	\newcommand{\hE}{\hat{E}}
	\newcommand{\hV}{\hat{V}}
	\newcommand{\calR}{\mathcal{R}}
	\newcommand{\hcalR}{\hat{\mathcal{R}}}
	\newcommand{\hid}{\hat{\mathds{1}}}
	\newcommand{\hsquare}{\hat{\square}}
	\renewcommand{\thefootnote}{\arabic{footnote}}
	\newcommand{\barlambda}{\overline{\lambda}}
	\setcounter{footnote}{0}
	\newcommand{\calL}{\cal{L}}

	\tableofcontents

	\section{Introduction}\label{sect:intro}
	
	Einstein's General Relativity (GR) provides a quantitatively good description of gravitational phenomena at stellar and planetary scales, while it does not yield the correct results if one considers galactic-scale dynamics, cosmological expansion unless exotic ingredients like dark matter and dark energy are introduced into dynamics. On the other hand,    the Ultra-Violet (UV) limit  suffers of the lack of a full quantum theory of gravity stemming out from GR.  During the last decades, many attempts have been made to  provide a novel explanation of gravitational physics which is devoid of these shortcomings. One of these attempts belongs to the family of the so-called Metric-Affine theories of Gravity (MAG), in which one extends  and modifies the geometrical setting of GR, i.e. the Riemannian geometry, aiming to find a new description of  physical phenomena in a more general framework.
	
	The basic idea of metric-affine theories is that the torsion-free and metricity conditions, assumed \emph{a priori} in GR, can actually be relaxed considering also  connection degrees of freedom (d.o.f.)  more general than the standard Levi-Civita one. Such a perspective was first put forward by  Palatini in his seminal work \cite{Palatini:1919ffw}, which shows that the metricity condition arises simply by computing the connection field equations of a dimension-two Lagrangian involving the generalized scalar curvature. A complementary approach was then proposed by Cartan, who first introduced the spin-connection and co-frame fields on purely geometrical footings \cite{Cartan:1923zea}: in this case,  the field equations of  spin-connection yield the vanishing torsion condition, and hence GR is once again obtained by going on-shell \footnote{Both these approaches to General Relativity are summarized in Chapter \S 14 of Ref. \cite{3122041}}. On the other hand, a generic metric-affine geometric theory includes both torsion and non-metricity, which can, in principle, give rise to further dynamical d.o.f.\ \cite{Hehl:1994ue}. However, these general geometries provide highly non-trivial dynamics, thus their  analysis has been limited to some simplified subclasses of models.
	
	The geometrical setup of MAGs can be introduced in two different ways. In the geometrical language by Ref. \cite{Hehl:1994ue} and other authors, an inner principal gauge bundle is postulated, whose structure group is $GL(4,\mathbb{R})$, and this internal space is glued to the tangent bundle of the manifold by introducing a frame bundle, i.e., a soldering form. In this case,  the affine connection of $GL(4,\mathbb{R})$ is the generalization of the spin-connection, the local metric in the internal space needs not be covariantly constant in the presence of non-metricity, and these two geometric quantities are treated as independent fields exactly like the soldering form. This first approach towards MAGs is usually called the {\it anholonomic one,} whose name is due to the fact that the partial derivatives,  defined in the internal space, do not satisfy Schwarz's theorem. On the other hand, one can project everything onto the coordinate or holonomic indices, which is usually thought of as a partial gauge-fixing \cite{Percacci:2023rbo}. Of course, the resulting geometric structure is less rich than the previous one since  the latter one only has the metric tensor and a general affine-connection as independent field variables. However, such a gauge-fixing procedure is usually applicable only in absence of fermionic d.o.f.\, and it is completely equivalent to the anholonomic one, if only bosons are present. Furthermore, the anholonomic approach allows for simpler manipulations of tensor equations and it is more easily employed in order to take into account the quantum aspects of the theory \cite{Melichev:2023lwj,Martini:2023apm,Melichev:2024hih,Melichev:2025hcg}.
	
	In the realm of metric-affine theories, one should also consider a notable class of them which stands on a yet different footing, whose models are usually referred to as \emph{teleparallel} theories \cite{Pereira}. These are given by those geometries in which the total curvature tensor of the manifold vanishes, whereas the torsion and/or the non-metricity are non-trivial. The first condition implies that parallel-transported vectors are always parallel to themselves, and it motivates the name itself of these theories. In this context, the simplest and most studied cases are those in which only the torsion or the non-metricity are non-zero, while in the general case both tensors can be non-trivial. In complete analogy with the general metric-affine setting, there are two geometric approaches for studying these theories. On the one hand, the local gauge-group formalism has been employed by Adak and his collaborators \cite{Adak:2005cd,Adak:2006rx,Adak:2008gd,Adak:2018vzk,Adak:2023ymc,Adak:2026mgj}, who first focused on symmetric teleparallel theories, i.e., those in which only the non-metricity is nonzero. On the other hand, much of the subsequent works on this subject have relied on the holonomic language. It is worth noticing that extensive efforts have been pursued to establish dynamical analogies and equivalences among MAGs,  GR and its extensions like $f(R)$ gravity. See e.g. \cite{BeltranJimenez:2017tkd,BeltranJimenez:2018vdo,BeltranJimenez:2019esp,BeltranJimenez:2019tme,Bahamonde:2021gfp,Capozziello:2022zzh,Capozziello:2023vne,Capozziello:2025hyw}. 
	
	Specifically, among the teleparallel theories of gravity there are two special classes of models  that deserve a wholly different focus, which are known as teleparellel equivalent theories. These are built by noticing that, since the total curvature vanishes, the Einstein-Hilbert action can be equivalently written in terms of a quadratic expression in either the torsion or the non-metricity \cite{Adak:2005cd,VanNieuwenhuizen:1981ae}, and the two resulting Lagrangians will provide alternative equivalent descriptions of GR, since the Lagrangians are identical up to boundary terms. However, the quantum counterparts of these  theories are not expected to be equivalent, since the renormalization process will produce, in general, counterterms that do not preserve the special ratio of the coefficients which appears in the tensor structures of the Lagrangians (see, e.g., \cite{Casadio:2021zai}). 
	
	Even though symmetric teleparallel theories have been hugely employed in recent years for phenomenological applications \cite{Cai}, there are still several open issues from a foundational viewpoint. In particular, there are several debates in the literature regarding the number of  propagating d.o.f.\ in $f(\mathbb{Q})$ theories \cite{Heisenberg, Golovnev1, Golovnev2}, while no such an issue arises in Riemannian geometry \cite{Kluson:2013hza,Barker:2025gon} or in non-teleparallel metric-affine theories \cite{Glavan:2023cuy}. This feature of a given classical model is usually addressed by studying its Hamiltonian formulation and cataloging its constraints along the lines of the so-called Dirac-Bergman algorithm \cite{Dirac:1950pj,Anderson:1951ta,Henneaux:1992ig}. In general covariant theories,  a further complication arises when one wants to pursue this kind of analysis, since one is forced to employ the Arnowitt-Deser-Misner (ADM)  formalism \cite{Arnowitt:1959ah}, which necessitates the introduction of a foliation of spacetime and the description of the four-dimensional geometry in terms of intrinsic and extrinsic tensors \cite{Poisson:2009pwt, Jackson}. Despite the efforts that have been put forward, there is still a lack of consensus regarding the number of dynamical d.o.f.\  in symmetric teleparallel theories. In fact,  the previous analysis has found different results and relied on a particular choice of gauge known as the \emph{coincident gauge} \cite{DAmbrosio:2023asf,Tomonari:2023wcs}, whose employment is not guaranteed to avoid hindering the whole analysis.

	Such a state of art motivates the present work, which aims at establishing a clear geometric footing for analyzing the Hamiltonian structure and the number of propagating d.o.f.\ of symmetric teleparallel theories. This is done by dealing with the simplest example of such theories, i.e., the Symmetric Teleparallel Equivalent of GR (STEGR). From a technical viewpoint, this goal is achieved by performing a complete and new study of the foliations of a spacetime manifold which bears non-metricity. Then, the teleparallel limit of the general results on non-metric foliations will then pick up the only nontrivial intrinsic and extrinsic tensors in \emph{any} teleparallel symmetric theory, prohibiting \emph{a priori} the possibility that some field variables have nontrivial conjugated momenta. Finally, the simple case of STEGR will be analyzed in detail, discussing the absence of boundary terms in the action and providing a proof that it shares the same number of propagating d.o.f.\ with GR. What happens in more general scenarios is out of the scope of the present paper, though the subsequent discussion provides a geometric framework which can be applied unambiguously to all symmetric teleparallel theories.
	
	The structure of the paper is as follows. In Sect.\ \ref{sect:intro-to-stegr}, the basic geometric definitions of symmetric teleparallel theories are given, whereas Sect.\ \ref{sect:constrained-systems} provides a very brief introduction to constrained Hamiltonian systems and the counting of degrees of freedom. Then, Sect. \ref{sect:foliations} starts from the fundamentals of foliations in Subsect.\ \ref{subsect:3+1-definitions} adapting the known results to the non-metric case, it introduces and classifies intrinsic and extrinsic non-metricity in Subsect.\ \ref{subsect:extrinsic-nonmetricity} and it constructs the $3+1$ decomposition of the manifold in Subsect.\ \ref{subsect:3+1}. The subsequent section \ref{sect:gauss-codazzi} is devoted to the derivation of the generalized Gauss-Codazzi relations in a non-metric geometry. In particular, in Subsect.\ \ref{subsect:generalized-gauss-codazzi}, these equations are derived in full detail, they are employed in Subsect.\ \ref{subsect:scalar-curvature-relation} to work out an expression for the scalar curvature and, through a different parameterization introduced in Subsect.\ \ref{subsect:new-set}, the teleparallel limit of these relations is studied, deducing which tensors can be dynamical in a generic teleparallel theory. Sect. \ref{sect:variational} is dedicated to the variational problem in non-metric theories, starting in Subsect.\ \ref{subsect:variation-palatini} with the Palatini action and dealing with the teleparallel case in Subsect.\ \ref{subsect:variation-telep-eq}. Finally, the aim of Sect. \ref{sect:hamiltonian-analyses} is to construct and analyze the Hamiltonian formulation of Palatini \ref{subsect:palatini-hamiltonian} and symmetric teleparallel equivalent theories. In particular, a detailed analysis of the latter is concluded by the proof that the number of propagating d.o.f.\ of STEGR and GR is the same. The paper closes with some remarks and pointing out the possible future directions in Sect.\ \ref{sect:conclusions}. Furthermore, two appendices supply some useful identities on integration by parts (App.\ \ref{app:sect:stokes}) and collects the intermediate steps of the variations performed in the main text (App.\ \ref{app:sect:variations}).

\section{The Symmetric teleparallel metric-affine gravity}\label{sect:intro-to-stegr}

In this section,  we briefly introduce the main ingredients that concur to the geometric formulation of symmetric teleparallel gravity. To do this, we choose to follow the holonomic approach, which is based on two field variables: the metric $g_{\mu\nu}$ and the affine-connection $\Gamma^\rho{}_{\nu\mu}$. In particular, given a contra-variant vector field $v^\rho$, its covariant derivative is given by
\begin{equation}
	\nabla_\mu v^\rho = \partial_\mu v^\rho + \Gamma^\rho{}_{\lambda\mu} v^\lambda \, .
\end{equation}
The affine-connection is required to be symmetric, i.e., its torsion must vanish
\begin{equation}
	\Gamma^\rho_{[\mu\nu]} = 0 \, .
\end{equation}
However, this connection is not meant to be metric-compatible, and it gives rise to the non-metricity tensor defined by
\begin{equation}\label{eq:non-metricity-def}
	Q_{\mu\alpha\beta} \equiv \nabla_\mu g_{\alpha\beta} \, .
\end{equation}
We are always free to split the connection into the sum of the metric-compatible Levi-Civita connection $\oGamma^\rho{}_{\lambda\mu}$ and a distortion tensor $L^\rho{}_{\lambda\mu}$ as
\begin{equation}\label{eq:gamma-split}
	\Gamma^\rho{}_{\lambda\mu} = \oGamma^\rho{}_{\lambda\mu} + L^\rho{}_{\lambda\mu} \, ,
\end{equation}
where all of the quantities entering this equation are symmetric in the lower indices. Consequently, the distortion tensor can be written in terms of the non-metricity as
\begin{equation}\label{eq:distortion-def}
	L^\alpha{}_{\beta\mu} = \frac{1}{2} \left( Q^\alpha{}_{\beta\mu} - Q_\beta{}^\alpha{}_\mu - Q_\mu{}^\alpha{}_\beta \right) \, ,
\end{equation}
which is manifestly symmetric in $\beta\leftrightarrow\mu$. The non-metricity has two independent traces, and it turns out to be convenient to introduce the following notation for dealing with them
\begin{eqnarray}\label{eq:non-metricity-vectors-def}
	Q_\mu \equiv Q_\mu{}^{\alpha}{}_{\alpha} \, , \qquad && \tilde{Q}_\mu \equiv Q_\alpha{}^\alpha{}_\mu \, .
\end{eqnarray}
In terms of these vectors, we can thus write down the two traces of the distortion tensor as
\begin{eqnarray}
	\tilde{L}_\nu \equiv L^\mu{}_{\mu\nu} = - \frac{1}{2} Q_\nu \, , \qquad && L^\mu \equiv L^{\mu\nu}{}_\nu = \frac{1}{2} Q^\mu - \tilde{Q}^\mu \, .
\end{eqnarray}
The full curvature tensor of the non-metric geometry is defined in the usual manner, and using Eq.\ \eqref{eq:distortion-def} it can be written as
\begin{equation}\label{eq:full-curvature-def}
	\begin{split}
		R^\rho{}_{\lambda\mu\nu} \equiv & \partial_\mu \Gamma^\rho{}_{\lambda\nu} - \partial_\nu \Gamma^\rho{}_{\lambda\mu} + \Gamma^\rho{}_{\alpha\mu} \Gamma^\alpha{}_{\lambda\nu} - \Gamma^\rho{}_{\alpha\nu} \Gamma^\alpha{}_{\lambda\mu} \\
		= & \mathring{R}^\rho{}_{\lambda\mu\nu} + \onabla_\mu L^\rho{}_{\lambda\nu} - \onabla_\nu L^\rho{}_{\lambda\mu} + L^\rho{}_{\alpha\mu} L^\alpha{}_{\lambda\nu} - L^\rho{}_{\alpha\nu} L^\alpha{}_{\lambda\mu} \, ,
	\end{split}
\end{equation}
where from now on we are denoting with a circle  objects  living in the purely Riemannian geometry. In the symmetric teleparallel geometry, the previous equation vanishes identically, therefore we can express the Riemann tensor and its traces in terms of the distortion tensor and its Levi-Civita covariant derivatives. In turn, we can then write the distortion tensor in terms of the non-metricity by using Eq.\ \eqref{eq:distortion-def}.

Due to the symmetry of the affine connection, the Bianchi identities take a very simple form in non-metric torsionless theories. Indeed, the first algebraic identity simply reads
\begin{equation}\label{eq:bianchi-1}
	R^\rho{}_{[\lambda\mu\nu]} = 0 \, .
\end{equation}
On the other hand, the differential Bianchi identity is written as
\begin{equation}\label{eq:bianchi-2}
	\nabla_{[\mu} R^\rho{}_{|\lambda|\nu\sigma]} = 0 \, .
\end{equation}
Both identities can be checked explicitly by performing the splitting of the affine connection Eq.\ \eqref{eq:gamma-split} and using the symmetry of the distortion tensor. 


\section{Constrained Hamiltonian systems}\label{sect:constrained-systems}

Let us now provide a brief introduction to the Hamiltonian description of constrained systems. We  shall introduce the definitions that are necessary to compute the number of independent d.o.f.\ of a given field theory. To avoid overburdening the notation, we choose to employ the language of a classical point-particle system to illustrate the relevant features. In particular, we will closely follow the treatment and notational conventions in \cite{Henneaux:1992ig}.

Let us consider a Lagrangian $L(\dot{q}^n(t),q^n(t))$, where the index $n$ runs over ${1,\dots,N}$. The canonical Hamiltonian $H$ is defined by the Legendre transformation
\begin{equation}\label{eq:def-hamiltonian}
	H = \dot{q}^n p_n - L \, ,
\end{equation}
where the momentum conjugated to the $n$-th coordinate is
\begin{equation}\label{eq:def-momentum}
	p_n \equiv \frac{\partial L}{\partial \dot{q}^n} \, .
\end{equation}
Then, the very definition of the Legendre transformation implies that the Hamiltonian is a function $H=H( q^n(t),p_n(t) )$. However, when building the Hamiltonian out of the Lagrangian, one may come across a possible issue, i.e., that we may not be able to invert the $\dot{q}^n$ as functions of the coordinates and momenta. This happens whenever the rank $K$ of the Hessian is $K<N$, and thus it has a vanishing determinant
\begin{equation}
	\det \frac{\partial^2 L}{\partial \dot{q}^n \partial \dot{q}^{n'}} = 0 \, ,
\end{equation}
When this is the case, we cannot simply perform the desired inversion. As a matter of fact, in order to span the degenerate subspace, one is required to introduce a set of expressions $\phi_m(q,p)$ which vanish as a direct consequence of the definition of the momenta Eq.\ \eqref{eq:def-momentum}, i.e.
\begin{eqnarray}\label{eq:def-primary-constraints}
	\phi_m(q,p) = 0 \, , \,\, {\rm where} \qquad && m=1,\dots,M \,\,\,\, {\rm and} \,\,\,\, M=N-K<N \, .
\end{eqnarray}
These relations are called \emph{primary constraints}, and are further required to have non-singular derivatives with respect to the canonical variables \cite{Henneaux:1992ig}. The primary constraints permit to span the $M$-dimensional degenerate subspace of phase space, and to have a well-defined Legendre transformation. These constraints can be enforced directly in the action functional by writing the following variational principle
\begin{eqnarray}\label{eq:modified-var-principle}
	\delta \,\, \int_{t_1}^{t_2} \left( \dot{q}^n p_n - H - u^m \phi_m \right) \, dt = 0 \, , && \qquad \delta q^n(t_1) = \delta q^n(t_2) = 0 \, .
\end{eqnarray} 
Here $u^m = u^m(q,p)$ are Lagrange multipliers. The equations of motion that follow from this variational principle are \cite{Henneaux:1992ig}
\begin{subequations}
	\begin{align}
		& \dot{q}^n = \frac{\partial H}{\partial p_n} + u^m \frac{\partial u^m}{\partial p_n} \, ;\\
		& \dot{p}_n = - \frac{\partial H}{\partial q^n} - u^m \frac{\partial u^m}{\partial q^n} \, ;\\
		& \phi_m(q,p) = 0 \, .
	\end{align}
\end{subequations}
At this point, it is natural to introduce the Poisson bracket, whose algebra determines the features of a specific model. This bracket is defined as usual as 
\begin{equation}\label{eq:poisson-bracket-def}
	\{ F, G \} = \frac{\partial F}{\partial q^n} \frac{\partial G}{\partial p_n} - \frac{\partial F}{\partial p_n} \frac{\partial G}{\partial q^n} \, .
\end{equation}
By this tool, we can express the same equations of motion of the Lagrangian formulation as
\begin{equation}\label{eq:poisson-bracket-eom}
	\dot{F} = \{ F, H\} + u^m \{F,\phi_m \} \, .
\end{equation}

In order to be consistent, a theory  must have that the primary constraints be satisfied at every time $t$. Therefore, we have to impose that their time derivatives vanish, which, using the bracket that we have just introduced, can be written as 
\begin{equation}\label{eq:def-secondary-constraint}
	\dot{\phi}_m = \{\phi_m,H\} + u^{m'} \{ \phi_m , \phi_{m'} \} = 0 \, .
\end{equation}
As we take into account all the $m$ in $\{1,\dots ,M\}$,  we encounter the following dichotomy \emph{after using the equations of motion}: the last equation is either an identity, or it imposes some further constraints. When the second option is found, the theory has \emph{secondary constraints}. The latter can be symbolically written as $\phi_k$, where $k=M+1,\dots,M+K=J$. Here $J$ is  the total number of primary plus secondary constraints, and $\phi_j$ runs over both kind of constraints.

Primary and secondary constraints force the theory to live on a subspace of the starting phase space. Thus, it may be the case that two quantities (in general tensor densities) are equal when all these constraints are satisfied but not off this constraint surface. When this is the case, we say that these two objects are \emph{weakly} equal, or we speak about a \emph{weak equality} between them, i.e.,
\begin{equation}\label{eq:weak-equality}
	F \approx G \,  \iff \, F - G = f^j(q,p) \phi_j \, .
\end{equation}

Endowed with such a new definition, we can further enforce the consistency of the theory in the following way. Indeed, since the secondary constraints themselves must be fulfilled at all times, we demand that the time derivative all of any constraint $\phi_j$ (primary and secondary) vanishes weakly, namely
\begin{equation}\label{eq:Lagr-mult-weak-condition}
	\{ \phi_j, H\} + u^m \{ \phi_j, \phi_m \} \approx 0 \, .
\end{equation}
This equation should be understood as dictating which conditions must be met by the Lagrange multipliers $u^\mu$. Thus, due to the relevance of this requirement for the consistency of the whole formalism, we briefly focus on the classes of solutions of Eq.\ \eqref{eq:Lagr-mult-weak-condition}. First, it is convenient to split the Lagrange multipliers as
\begin{equation}
	u^m = U^m + V^m  \, ,
\end{equation}
where $V^m$ is the most general solution of the homogeneous equation, whereas $U^m$ is one particular solution of the inhomogeneous one. Thus, in the first case we have
\begin{equation}
	V^m \{ \phi_j,\phi_m \} \approx 0 \, .
\end{equation}
In general, there will be $a=1,\dots,A$ different independent solutions $V_a{}^m$ of the previous equation. By assuming that the matrix $\{ \phi_j,\phi_m \}$ has constant rank throughout phase space, the number $A$ is guaranteed to be the same for all $q$'s and $p$'s. Therefore, we may write the Lagrange multiplier as
\begin{equation}
	u^m \approx U^m + v^a V_a{}^m \, ,
\end{equation}
where $v^a$ are \emph{completely arbitrary} coefficients.

In order to proceed further, we will assume that the constraints $\phi_j=0$ can be put in an irreducible form. Even though it is not strictly necessary to do so, this assumption simplifies the subsequent analysis while retaining its relevant features. The next step is that of defining a total Hamiltonian as
\begin{equation}
	H_T = H' + v^a \phi_a = H + U^m \phi_m + v^a \phi_a \, ,
\end{equation}
whose definition allows us to write the equations of motion in a simple way as
\begin{equation}
	\dot{F} \approx \{F,H_T\} \, .
\end{equation}
The classical dynamics is then dictated by the full Hamiltonian, in which the set of completely arbitrary Lagrange multipliers is tightly connected to the presence of symmetries in the Lagrangian description of the same system \cite{Henneaux:1992ig}.

Apart from the primary and secondary constraints, there is yet another dichotomy which provides a novel classification of phase space functions. Let us consider a function of the phase space variables $F(q,p)$: its Poisson brackets with the primary and secondary constraints $\phi_j$ can either vanish or not on the constrained phase space. In the first case, we say that the function is of \emph{first-class}
\begin{equation}
	\{F,\phi_j\} \approx 0 \, ,
\end{equation}
whereas, in the latter one, we dub it a \emph{second-class} function. 

Equipped with the previous definitions, we are now able to state the standard result for the number of independent d.o.f.\  in a given constrained system. To this end, we consider the total number of canonical variables $\#(c.v.)$, the number of first-class constraints $\#(f.c.c.)$ and that of the independent original second-class constraints $\#(o.s.c.c.)$. In terms of these integers,  the number of propagating d.o.f.\  reads \cite{Henneaux:1992ig}
\begin{equation}
	\# (d.o.f.) = \frac{ \,\#(c.v.)\, - \, \#(o.s.c.c.) \, }{2} - \#(f.c.c.) \, .
\end{equation}
This result will be employed in Subect.\ \ref{subsect:stegr-hamiltonian} to prove that the number of d.o.f.\  of the teleparallel equivalent of GR is the same of GR itself.

\section{Foliations and hypersurfaces}\label{sect:foliations}

In order to derive the number of dynamical d.o.f.\ using the Dirac-Bergman algorithm we must have a clear notion of time and of the temporal derivatives of tensors. To introduce such notions in a generally covariant theory we are required to work out a foliation of the spacetime manifold. Consequently, any four-dimensional tensor is expressed through a collection of three-dimensional tensors with rank less than or equal to the original one, where the highest rank part describes the intrinsic  geometry, whereas the extrinsic one is parametrized by lower rank tensors.

\subsection{Basic geometric definitions}\label{subsect:3+1-definitions}

In this subsection we introduce the necessary ingredients for performing a $3+1$ decomposition of theories with non-metricity. To this end we follow the pedagogical presentation of \cite{Poisson:2009pwt}, generalizing some geometric notions whenever it is necessary. Here we shall be as general as possible, performing all the calculation for a generic non-null family of hypersurfaces foliating spacetime, and all our formulae will be valid in \emph{any} spacetime dimension $d$. The applications of the present results to the actual time-space splitting of the spacetime manifold will be carried out in the subsequent section.

The geometric problem of foliating spacetime amounts to finding a globally defined vector field $n^\mu$, such that this vector field is everywhere orthogonal to the three-dimensional hypersurpaces $\Sigma$, such that the foliation of spacetime induced by this decomposition is smooth. It is customary to normalize the vector $n^\mu$ such that
\begin{eqnarray}\label{eq:norm-n}
	n^\mu n_\mu = \epsilon \, , \quad {\rm where} \quad \epsilon = -1 \,\,\, {\rm for} \,\,\, \Sigma \,\,\, {\rm spacelike,} && \quad \epsilon = +1 \,\,\, {\rm for} \,\,\, \Sigma \,\,\, {\rm timelike} \, .
\end{eqnarray}
Let us parametrize the hypersurface $\Sigma$ by choosing some coordinates $y^a$ on it, with $a=1,2,3$. Then, if $x^\mu$ are the spacetime coordinates we have that the three vector fields
\begin{eqnarray}\label{eq:def-basis-vectors}
	e^\mu{}_a = \frac{\partial x^\mu}{ \partial y^a} \, , && \qquad  e^\mu{}_a n_\mu = 0 \, ,
\end{eqnarray}
belong to the tangent bundle of the submanifold $\Sigma$, and the \emph{induced metric} on this hypersurface is
\begin{equation}\label{eq:def-induced-metric}
	\begin{split}
		d s^2_\Sigma = & \, g_{\mu\nu} \frac{\partial x^\mu}{\partial y^a} \frac{\partial x^\nu}{\partial y^b} dy^a dy^b \\
		\equiv & \, h_{ab} d y^a d y^b \, ,
	\end{split}
\end{equation}
and it is also referred to as the \emph{first fundamental form}. Accordingly, the completeness relation for the four-dimensional metric tensor reads
\begin{equation}\label{eq:completeness-metric}
	g_{\mu\nu} = \epsilon \, n_\mu n_\nu + e_\mu{}^a e_\nu{}^b h_{ab} \, ,
\end{equation}

Let us now consider a $1$-form $A_\mu$ which belongs to the co-tangent bundle of the submanifold $\Sigma$, i.e.,
\begin{equation}
	A_\mu n^\mu = 0 \, .
\end{equation}
We are looking a notion of covariant differentiation and parallel displacement on the hypersurface $\Sigma$. Whenever $A_\mu$ satisfies the previous condition we can define the \emph{intrinsic covariant derivative} $D_a$ as
\begin{equation}\label{eq:def-3-covd}
	\begin{split}
		D_a A_b \equiv & \, e^\alpha{}_a e^\beta{}_b \nabla_\beta A_\alpha\\
		= & \, \nabla_\beta \left( e^\alpha{}_a A_\alpha  \right) e^\beta{}_b - e^\beta{}_b A_\alpha \nabla_\beta e^\alpha{}_a\\
		= & \, \partial_b A_a - \Gamma^c{}_{ab} A_c \, .
	\end{split}
\end{equation}
The \emph{intrinsic affine-connection} is implicitly defined by the previous equation, and its explicit form is
\begin{equation}\label{eq:Gamma3}
	\Gamma^c{}_{ab} = \oGamma^c{}_{ab} + {L^{(3)}}{}^c{}_{ab} = e^\beta{}_b e^{\rho c} \onabla_\beta e_{\rho a} + L^\rho{}_{\nu\mu} e^\nu{}_a e^\mu{}_b e_\rho{}^c \, ,
\end{equation}
where $\oGamma^c{}_{ab}$ is the intrinsic Levi-Civita connection. By acting with this intrinsic covariant derivative on the first fundamental form and using Eq.\ \eqref{eq:def-3-covd} we find that the intrinsic non-metricity is indeed the non-metricity of the induced metric $h_{ab}$, i.e., 
\begin{equation}\label{eq:intrinsic-non-metricity}
	D_c h_{ab} = Q^{(3)}{}_{cab} \, .
\end{equation}
This is the first step towards the construction of intrinsic and extrinsic geometric quantities in the presence of non-metricity. The second one is provided by noticing that the covariant of a contra-variant vector $A^\mu$ along the $b$ direction of the hypersurface $\Sigma$ can be written as
\begin{equation}
	\begin{split}
		e^\nu{}_b \nabla_\nu A^\mu = e^\mu{}_c D_b A^c - \epsilon \, n^\mu K_{cb} A^c \, ,
	\end{split}
\end{equation}
where we have introduced the \emph{extrinsic curvature} or \emph{second fundamental form}
\begin{equation}\label{eq:extrinsic-curvature}
	\begin{split}
		K_{ab} \equiv & e^\mu{}_a e^\nu{}_b \nabla_\nu n_\mu \\
		= & \oK_{ab} - L^\lambda{}_{\mu\nu} n_\lambda e^\mu{}_a e^\nu{}_b \, . 
	\end{split}
\end{equation}
An analogous calculation shows that the generalization of the Gauss-Weingarten equation reads
\begin{equation}\label{eq:parallel-transport-basis-vectors}
	e^\nu{}_a \nabla_\nu e^\mu{}_b = e^\mu{}_c \Gamma^c{}_{ba} - \epsilon \, K_{ba} n^\mu \, .
\end{equation}
Then, using the fact that the basis vectors $e^\mu{}_a$ are transported one along the other, we deduce the symmetry of the extrinsic curvature
\begin{equation}
	K_{ab} = K_{(ab)} \, .
\end{equation}
Finally, we observe that this tensor can also be expressed in the following geometric way
\begin{equation}\label{eq:extrinsic-curvature-geometric}
	K_{ab} = \frac{1}{2} \left( \pounds_n g_{\mu\nu} \right) e^\mu{}_a e^\nu{}_b - L^\lambda{}_{\mu\nu} n_\lambda e^\mu{}_a e^\nu{}_b \, .
\end{equation}
Up to now we have merely generalized the geometric notions that are known in the $3+1$ decomposition of a pseudo-Riemannian geometry. However, the complicated effect of having a non-trivial non-metricity is the need to define more extrinsic quantities. We shall generally refer to these objects as \emph{extrinsic non-metricity}, and they are expressed as tensors, vectors and scalars belonging to the three-dimensional hypersurface $\Sigma$. 

\subsection{Intrinsic and extrinsic non-metricity}\label{subsect:extrinsic-nonmetricity}

While in a (pseudo-)Riemannian geometry one only needs to define the extrinsic curvature $\oK_{ab}$ and the intrinsic one, more tensors are required in the presence of non-metricity. In analogy with the purely Riemannian case the extrinsic tensors can be defined in terms of the covariant derivatives of the basis co-vectors and shift vectors, and they are given by two mixed-symmetry rank-$2$ tensors, a symmetric $2$-tensor, three vectors and a scalar. We will first define these tensors, and then show how they enter the generalizations of the Gauss-Codazzi equations for the non-metricity in the next section \ref{sect:gauss-codazzi}.

Let us start by defining the \emph{intrinsic non-metricity}, i.e.,
\begin{equation}
	Q^{(3)}{}_{cab} \equiv e^\mu{}_a e^\nu{}_b e^\rho{}_c \nabla_\rho g_{\mu\nu} \, .
\end{equation}
It is easy to show that this expression adheres with the one given previously, see Eq.\ \eqref{eq:intrinsic-non-metricity}. There are two rank-$2$ tensors that can be defined by covariantly differentiating the basis co-vectors $e_\mu{}^a$ and $e_{\mu a}$ along the shift vector, i.e.,
\begin{subequations}\label{eqs:extrinsic-non-metricity-tensors-n-der}
	\begin{align}
		& \Psi^{ab} \equiv e^{\mu a} n^\rho \nabla_\rho e_\mu{}^b \, ;\\
		& \Lambda_{ab} \equiv e^\mu{}_a n^\rho \nabla_\rho e_{\mu b} \, .
	\end{align}
\end{subequations}
Both tensor structures are needed to the non-metricity of the theory. On the other hand, as we are going to see in a while, only one rank-two tensor needs to be written as the covariant derivative along the basis vectors
\begin{equation}\label{eqs:extrinsic-non-metricity-tensor-e-der}
	\Phi^{ab} \equiv n^\mu e^{\rho a} \nabla_\rho e_\mu{}^b \, .
\end{equation}
Notice that the latter expression needs not be symmetric. Finally, one manifestly symmetric tensor can be defined as the covariant derivative along the shift vector of the induced metric
\begin{equation}\label{eqs:extrinsic-non-metricity-tensor-n-der-h}
	\Xi_{ab} \equiv n^\rho h_a{}^c h_b{}^d \partial_\rho h_{cd} \, .
\end{equation}
There exists a linear relation among the two tensors Eqs.\ \eqref{eqs:extrinsic-non-metricity-tensors-n-der} and $\Xi_{ab}$, i.e.,
\begin{equation}
	\Xi_{ab} = \Lambda_{(ab)} - \Psi_{(ab)} \, .
\end{equation}
Thus, one is free to choose a basis in this symmetric subspace of the extrinsic non-metricity tensors.

The \emph{extrinsic non-metricity vectors} can be defined in a similar fashion as
\begin{subequations}\label{eqs:extrinsic-non-metricity-vectors}
	\begin{align}
		& \theta_a \equiv 2 n^\mu e^\rho{}_a \nabla_\rho n_\mu \, ;\\
		& \kappa_a \equiv e^\mu{}_a n^\rho \nabla_\rho n_\mu \, ;\\
		& \lambda^a \equiv n^\rho n^\mu \nabla_\rho e_\mu{}^a \, .	
	\end{align}
\end{subequations}
Notice that in all of these three expressions the differentiated $3$-dimensional indices of the basis co-vectors are contra-variant. Finally, there is only one scalar, whose expression is
\begin{equation}\label{eq:extrinsic-non-metricity-scalar}
	\alpha \equiv n^\nu n^\nu \nabla_\nu n_\mu  \, .	
\end{equation}
By employing the previous relations for the covariant derivative of the shift vector we can write the completeness relation
\begin{equation}\label{eq:completeness-rel-covd-n}
	\nabla_\mu n_\nu = \, \alpha \, n_\mu n_\nu + \epsilon\, n_\mu e_\nu{}^a \kappa_a + \frac{\epsilon}{2} n_\nu e_\mu{}^a \theta_a + e_\mu{}^a e_\nu{}^b K_{ab} \, .
\end{equation}
We observe that in the Riemannian limit we have $\mathring{\Phi}_{ab} = - \oK_{ab}$, and $\mathring{\lambda}_a = - \mathring{\kappa}_a$, whereas all the other tensors vanish.

Having defined these tensors, we can now write down the projections of the four-dimensional non-metricity tensor in a simple way. Indeed, by starting with rank-$2$ expressions we find the following identities for the \emph{extrinsic non-metricity tensors}
\begin{eqnarray}
	Q_{\rho\mu\nu} n^\rho e^\mu{}_a e^\nu{}_b = \Xi_{ab} + 2 \Psi_{(ab)} \, , && \qquad Q_{\rho\mu\nu} n^\mu e^\nu{}_a e^\rho{}_b = K_{ab} + \Phi_{ba} \, .
\end{eqnarray}
Turning to the vector contractions, we derive the subsequent relations for the \emph{extrinsic non-metricity vectors}
\begin{eqnarray}
	e^\rho{}_a n^\mu n^\nu Q_{\rho\mu\nu} = \theta_a \, , && \qquad n^\mu n^\nu Q_{\mu\nu}{}^\lambda e_{\lambda a} = \kappa_a + \lambda_a \, .
\end{eqnarray}
Ultimately, the \emph{extrinsic non-metricity scalar} is given by
\begin{equation}
	Q_{\rho\mu\nu} n^\rho n^\mu n^\nu = 2 \alpha \, .
\end{equation}
These projections of the non-metricity onto $n^\mu$ and $e^\mu{}_a$ provide the necessary ingredients to work express every four-dimensional scalar as a linear combination of three-dimensional ones, as well as a completeness relation for the four-dimensional non-metricity itself, i.e.,
\begin{align}\label{eq:completeness-non-metricity}
	Q^{\rho\mu\nu} = \,\,& 2 \epsilon \, \alpha n^\rho n^\mu n^\nu + n^\rho n^\mu e^{\nu a} \left( \kappa_a + \lambda_a \right) + n^\rho n^\nu e^{\mu a} \left( \kappa_a + \lambda_a \right) + n^\mu n^\nu e^{\rho a} \theta_a \\\nonumber
	& + \epsilon \left[ \left( n^\mu e^{\rho a} e^{\nu b} \left( K_{ab} + \Phi_{ab} \right) + \left( {\footnotesize \mu} \leftrightarrow {\footnotesize \nu} \right) \right) + n^\rho e^{\mu a} e^{\nu b} \left( \Xi_{ab} + 2 \Psi_{(ab)} \right) \right] + e^{\rho c} e^{\mu a} e^{\nu b} Q^{(3)}{}_{cab } \, .
\end{align}
From this equation we deduce that the non-metricity is independent of the antisymmetric part of $\Psi_{ab}$. From the previous expression we can derive four-dimensional relations for the two vectors and two tensors that parametrize the intrinsic non-metricity. In the former case we find
\begin{eqnarray}
	\begin{split}
		Q_\rho{}^{\nu\alpha} n^\rho n_\alpha = 2 \epsilon \, \alpha n^\nu + \left( \kappa_a + \lambda_a \right) e^{\nu a} \, , && \qquad Q_\rho{}^{\mu\nu} n_\mu n_\nu = 2 \epsilon \, \alpha n_\rho + \theta_a e_\rho{}^a \, .
	\end{split}
\end{eqnarray}
The expressions of the tensor parts are more complicate, and the manifestly symmetric one is
\begin{align}
	Q_\rho{}^{\mu\nu} n^\rho = 2 n^\mu n^\nu \alpha + \epsilon \left[ e^{\mu a} n^\nu \left( \kappa_a + \lambda_a \right) + e^{\nu a} n^\mu \left( \kappa_a + \lambda_a \right) \right] + e^{\mu a } e^{\nu b} \left( \Xi_{ab} + 2 \Psi_{(ab)} \right) \, ,
\end{align}
whereas the second one is
\begin{equation}
	Q^{\rho\mu\nu} n_\nu = 2 n^\rho n^\mu \alpha + \epsilon \left[ n^\rho e^{\mu a} \left( \kappa_a + \lambda_a \right) + n^\mu e^{\rho a} \theta_a \right] + e^{\rho a} e^{\mu b} \left( K_{ab} + \Phi_{ab} \right) \, .
\end{equation}

Before delving into the details of this calculation, we write the completeness relations for the two algebraically independent traces of the non-metricity. By tracing the first two indices we find
\begin{equation}\label{eq:completeness-first-trace}
	\begin{split}
		Q_\mu{}^{\mu\lambda} = n^\lambda \left( 2 \alpha  + \epsilon \, \left( K^a{}_a + \Phi^a{}_a \right) \right) + e^{\lambda a} \left( \epsilon \left( \kappa_a + \lambda_a \right)  + Q^{(3)}{}_b{}^b{}_a  \right) \, ,
	\end{split}
\end{equation}
whereas taking the trace on the last two indices yields
\begin{equation}\label{eq:completeness-second-trace}
	Q_\rho{}^\mu{}_\mu = n_\rho \left( 2 \alpha + \epsilon \left( \Xi^a{}_a + 2 \Psi^a{}_a \right) \right) + e_\rho{}^a \left( \epsilon \, \theta_a + Q^{(3)}_a{}^b{}_b \right) \, .
\end{equation}

Before delving into the details of the Gauss-Codazzi equations, we observe that the Riemannian limit is given by
\begin{eqnarray}\label{eq:riemannian-limit}
	\mathring{\Phi}_{ab} = - \mathring{K}_{ab} \, , && \quad \mathring{\kappa}_a = - \mathring{\lambda}_a = a_a \, , \qquad \mathring{\theta}_a =0 \, , \qquad \mathring{\Xi}_{ab} = \mathring{\Psi}_{(ab)}=0 \, , \qquad \mathring{\alpha} = 0 \, .
\end{eqnarray}
In a Riemannian geometry the $a_a$ vector is known as the \emph{acceleration vector}, and it enters scalar expressions in a nontrivial way, in particular when one considers higher-derivative theories \cite{Kluson:2013hza}. The previous relations will be used at the end of the subsequent calculations in order to check the consistency with known results.

\subsection{$3+1$ space-time decomposition}\label{subsect:3+1}

In this subsection we shall apply the machinery of foliations that we have introduced above to study the space-time decomposition of the manifold.

First of all, let us postulate the existence of a scalar field $t(x^\mu)$ such that $t=c$ describes a family of spacelike hypersurfaces $\Sigma_t$ which are not intersecting with each other. On $\Sigma_t$ we use coordinates $y^a$, which are related to the coordinates at on a different time slice by requiring the existence of a congruence of curves $\gamma$ that intersect $\Sigma_t$. Thus, points $p \in \Sigma_t$ and $p' \in \Sigma_{t'}$ have the same coordinates $y^a$ if and only if there exists a curve which joins them. Then, $t$ can be employed as a parameter on such curves, and the future-directed timelike vector field $t^\mu$ is tangent to the congruence as satisfies
\begin{equation}\label{eq:def-time-vector}
	t^\mu \partial_\mu t = 0
\end{equation}
needs not to be orthogonal to the spacelike hypersurfaces $\Sigma_t$, but it must have a non-zero component along the outward future directed timelike $n^\mu$ direction. Therefore, it is written as
\begin{equation}\label{eq:time-vector}
	t^\mu = N n^\mu + e^\mu{}_a N^a \, ,
\end{equation}
where the lapse function $N$ ensures the correct normalization, while the shift vector $N_a$ parametrizes how much $t^\mu$ differs from $n^\mu$. The differential $dx^\mu$ can then be expanded as
\begin{equation}
	dx^\mu = t^\mu dt + e^\mu{}_a dy^a = \left( N n^\mu + e^\mu{}_a N^a \right) dt + e^\mu{}_a dy^a \, .
\end{equation}
Finally, from the previous equation we can now derive space-time decomposition of the metric tensor, which reads
\begin{equation}\label{eq:3+1-metric}
	g_{tt} = - N^2 + N^a N_a \, , \qquad g_{ta} = N_a \, , \qquad g_{ab} = h_{ab} \, .
\end{equation}

The $3+1$ decomposition of the non-metricity can be worked out by contracting the indices with the basis $1$-forms, which is a coordinate invariant expression
\begin{equation}
	Q_{\rho\mu\nu} dx^\rho \otimes \left( d x^\mu \otimes_S d x^\nu \right) = Q_{\rho\mu\nu} \left( t^\rho dt + e^\rho{}_c dy^c \right) \otimes \left( \left( t^\mu dt + e^\mu{}_a dy^a \right) \otimes_S \left( t^\nu dt + e^\nu{}_b dy^b \right) \right)  \, ,
\end{equation}
where $\otimes_S$ is the symmetric tensor product. From this equation we read out the components of the non-metricity tensor in the $(t,y^a)$ coordinates. The three-scalar component is
\begin{equation}
	\begin{split}
		Q_{ttt} = & \, 2 \alpha^3 N^3 + N^2 \left( \theta_a + 2 \left( \kappa_a + \lambda_a \right) \right) N^a \\\
		& + N \left( \Xi_{ab} + 2 \Psi_{(ab)} + 2\left(K_{ab} + \Phi_{ab}  \right) \right) N^a N^b + Q^{(3)}{}_{cab} N^c N^a N^b \, .
	\end{split}
\end{equation}
There are two three-vector components, the first of which reads
\begin{equation}
	Q_{tat} = N^2 \left( \kappa_a + \lambda_a \right) + N \left( \Xi_{ab} + 2 \Psi_{(ab)} + K_{ab} + \Phi_{ab} \right) N^b + Q^{(3)}{}_{cab} N^c N^b \, ,
\end{equation}
whereas the other one takes the following form
\begin{equation}
	Q_{ctt} = N^2 \theta_c + 2 N \left( K_{ca} + \Phi_{ca} \right) N^a + Q^{(3)}{}_{cab} N^a N^b \, .
\end{equation}
We have two rank-$2$ three-tensors: the first is asymmetric and is given by
\begin{equation}
	Q_{cat} = N \left( K_{ca} + \Phi_{ca} \right) + Q^{(3)}{}_{cab} N^b \, ,
\end{equation}
while the second one is symmetric and its expression is
\begin{equation}
	Q_{tab} = N \left( \Xi_{ab} + 2 \Psi_{(ab)} \right) + Q^{(3)}{}_{cab} N^c \, .
\end{equation}
Finally, the rank-$3$ three-tensor part is exactly the intrinsic non-metricity of $\Sigma_t$.
\begin{equation}
	Q_{cab} = Q^{(3)}{}_{cab} \, .
\end{equation}

\section{Gauss-Codazzi equations and beyond}\label{sect:gauss-codazzi}

The most important issue that is posed by having a foliation of spacetime is how to express the four-dimensional curvature in terms of the three-dimensional one and of the extrinsic curvature. In Riemannian geometry the celebrated Gauss-Codazzi equations provide such a projection, thus, we now focus on generalizing these results to the case of a non-metric geometry. In order to do so, we start by following the derivation of \cite{Poisson:2009pwt} of the Gauss-Codazzi equations, and then we discuss the need to work out some further relations.

Let us commence by acting with $e^\rho{}_c \nabla_\rho$ on both sides of Eq.\ \eqref{eq:parallel-transport-basis-vectors}, which yields
\begin{equation}
	e^\rho{}_c \nabla_\rho \left( e^\nu{}_b \nabla_\nu e^\mu{}_a \right) = e^\rho{}_c \nabla_\rho \left( \Gamma^d{}_{ba} e^\mu{}_d - \epsilon \, K_{ba} n^\mu \right) \, .
\end{equation}
Then, manipulate this identity solving for $e^\rho{}_c e^\nu{}_b \nabla_\rho \nabla_\nu e^\mu{}_a$, we antisymmetrize in $b\leftrightarrow c$, and we make use of the definition of the three-dimensional covariant derivative Eq.\ \eqref{eq:def-3-covd}, thus finding
\begin{equation}\label{eq:gauss-codazzi-0}
		R^\mu{}_{\lambda\rho\nu} e^\lambda{}_a e^\rho{}_c e^\nu{}_b = \, e^\mu{}_d \left[ R^{(3)}{}^d{}_{acb} + 2 \epsilon \, K_{[ba} \Phi_{c]}{}^d \right] + \epsilon \, n^\mu \left( 2 D_{[b} K_{c]a} - \epsilon \, \theta_{[b} K_{c]a} \right)  \, .
\end{equation} 
From this equation we readily derive the Gauss-Codazzi equations by contraction with $e_\mu{}^f$ and $n_\mu$. In the first case we find the generalization of the \emph{Gauss relation}, i.e.,
\begin{empheq}[box={\mymath[colback=white!30, sharp corners]}]{equation}\label{eq:gauss-relation}
		R^\mu{}_{\lambda\rho\nu} e_\mu{}^d e^\lambda{}_a e^\rho{}_c e^\nu{}_b = R^{(3)}{}^d{}_{acb} + 2 \epsilon \, K_{a[b} \Phi_{c]}{}^d \, .
\end{empheq}
On the other hand, by contracting Eq.\ \eqref{eq:gauss-codazzi-0} with the normal vector we obtain the \emph{first generalized Codazzi relation}
\begin{empheq}[box={\mymath[colback=white!30, sharp corners]}]{equation}\label{eq:first-generalized-codazzi}
	R^\mu{}_{\lambda\rho\nu} n_\mu e^\lambda{}_a e^\rho{}_c e^\nu{}_b = D_b K_{ca} - D_c K_{ba} + \frac{\epsilon}{2} \left( \theta_c K_{ba} - \theta_b K_{ca} \right)  \, .
\end{empheq}
In the Riemannian case there is only one further nontrivial contraction of the Riemann tensor which must be taken into account, i.e., that which is quadratic in the orthogonal vector $n_\mu$. However, in a non-metric geometry the full Riemann tensor does not share the same symmetries of $\mathring{R}^\rho{}_{\lambda\mu\nu}$, so more contractions must be considered.

\subsection{Generalized Gauss-Codazzi-Ricci relations}\label{subsect:generalized-gauss-codazzi}

In order to examine the new relations that are not present in the Riemannian case we now employ a manifestly four-dimensional formalism. The latter is set by writing the metric tensor as
\begin{equation}
	g_{\mu\nu} = \epsilon \, n_\mu n_\nu + h_{\mu\nu} \, ,
\end{equation}
and by defining the induced three-dimensional covariant derivative acting on a vector $A^\mu$ orthogonal to $n_\mu$ as
\begin{equation}
	D_\mu A^\nu \equiv h^\rho{}_\mu h_\lambda{}^\nu \nabla_\rho A^\lambda \, .
\end{equation}
Due to the non-metric nature of the present geometry one must be careful at manipulating the indices. In particular, the definitions of the extrinsic curvature and non-metricity given above extend naturally by keeping the positions of the indices unchanged also in the present formalism. This allows us to pass from one framework to the other seamlessly.

Of particular importance for the present section is the comparison with the work of \cite{Kluson:2013hza}, in which the decomposition of the Riemann tensor and of its traces is studied extensively in order to analyze the Hamiltonian structure of higher-derivative theories. As we have already mentioned above, the symmetries of the Riemann tensor tell us that there is only one independent contraction with two normal vectors and two induced metrics, which is referred to as \emph{Ricci equation}. On the other hand, in the present case we have three such equations, where the first two relations yield the Riemannian result in the appropriate limit, whereas the last one is completely new and is due to the asymmetry of the full curvature tensor in its $GL(4)$-valued indices.

\subsubsection{Ricci relations}\label{subsubsect:ricci-relations}

The Ricci relations are found by contracting twice the full curvature tensor with the normal vector $n_\mu$. As we have already noticed, this can be done in three different ways when the geometry has non-metricity. In particular, the first two Ricci relations are found by contracting one of the last two indices and one of the first two $n_\mu$. Due to the asymmetry of the curvature tensor different results are found by picking up the first or the second $GL(4)$-valued indices, resulting in the following \emph{first and second generalized Ricci relations}
\begin{empheq}[box={\mymath[colback=white!30, sharp corners]}]{equation}\label{eq:ricci-relation-1}
	R^\mu{}_{\alpha\beta\nu} h^\alpha{}_\lambda n_\mu n^\beta h^\nu{}_\rho = D_\rho \kappa_\lambda - \epsilon \, \kappa_\rho \kappa_\lambda + \epsilon \, \alpha K_{\lambda\rho} - K_{\rho\mu} \Phi_\lambda{}^\mu - \pounds_n K_{\lambda\rho} \, ,
\end{empheq}
and
\begin{empheq}[box={\mymath[colback=white!30, sharp corners]}]{equation}\label{eq:ricci-relation-2}
R^\mu{}_{\alpha\nu\beta} n^\alpha n^\beta h^\sigma{}_\mu h^\nu{}_\rho = - D_\rho \lambda^\sigma + \epsilon \, \lambda^\sigma \left( \kappa_\rho - \theta_\rho \right) + \epsilon \, \alpha \Phi_\rho{}^\sigma - \Phi_\mu{}^\sigma \Phi_\rho{}^\mu + h^\sigma{}_\mu \pounds_n \Phi_\rho{}^\mu \, .
\end{empheq}
A different nontrivial result is found when we contract the first two indices of the curvature tensor with the normal vector, yielding the \emph{third generalized Ricci relation}
\begin{empheq}[box={\mymath[colback=white!30, sharp corners]}]{equation}\label{eq:ricci-relation-last}
	R^\mu{}_{\lambda\rho\nu} n^\lambda n_\mu h^\rho{}_\alpha h^\nu{}_\beta = 2 K_{\mu[\alpha} \Phi_{\beta]}{}^\mu - D_{[\alpha} \theta_{\beta]} \, ,
\end{empheq}
which vanishes in the metric limit. Indeed, by employing the Riemannian limit of the extrinsic tensors which is summarized in Eq.\ \eqref{eq:riemannian-limit} one can easily check that the right-hand side is zero identically.

\subsubsection{Rank-one relation}\label{subsubsect:rank-one-relation}

As we have already observed, the presence of non-metricity gives rise to yet more complicated features with respect to the Riemannian case. Indeed, there is now a nontrivial contraction of the full curvature tensor with three normal vectors, which results in a rank-one tensor expression. For this reason, we shall dub this equation as \emph{rank-one relation}
\begin{empheq}[box={\mymath[colback=white!30, sharp corners]}]{equation}\label{eq:rank-one-relation}
	R^\mu{}_{\lambda\rho\nu} n^\lambda n_\mu n^\rho h^\nu{}_\alpha = D_\alpha \alpha + \epsilon \, \alpha \left( \tfrac{1}{2} \theta_\alpha - \kappa_\alpha \right) + \kappa^\mu \Phi_{\alpha\mu} - \lambda^\mu K_{\alpha\mu} - \tfrac{1}{2} \pounds_n \theta_\alpha  \, .
\end{empheq}
Again, we see that by using the Riemannian limit (see Eq.\ \eqref{eq:riemannian-limit}) we obtain a trivial identity, since both $\alpha$ and $\theta$ vanish in this limit, whereas the remaining expression exactly cancels out.

\subsubsection{Codazzi relations}\label{subsubsect:codazzi-relations}

Let us now turn to the analysis of the rank-three relations. The first equation of this kind is the \emph{first generalized Codazzi relation} that we have derived before, whose expression in the present formalism is
\begin{empheq}[box={\mymath[colback=white!30, sharp corners]}]{equation}\label{eq:first-generalized-codazzi-2}
	R^\mu{}_{\lambda\rho\nu} n_\mu h^\lambda{}_\alpha h^\rho{}_\beta h^\nu{}_\sigma = 2 D_{[\sigma} K_{\beta]\alpha} + \epsilon \, \theta_{[\beta} K_{\sigma]\alpha}  \, .
\end{empheq}
A very similar relation is found by contracting the normal vector with the second index of the full curvature instead of the first one, resulting in the \emph{second generalized Codazzi relation}
\begin{empheq}[box={\mymath[colback=white!30, sharp corners]}]{equation}\label{eq:second-generalized-codazzi}
	R^\mu{}_{\lambda\rho\nu} n^\lambda h_{\mu\sigma} h^\rho{}_\alpha h^\nu{}_\beta = - 2 D_{[\alpha} \Phi_{\beta]}{}^\sigma + \epsilon \, \theta_{[\beta} \Phi_{\alpha]}{}^\sigma  \, .
\end{empheq}
Clearly, both of the two previous Codazzi relations correctly reproduce the known result in Riemannian limit once we employ Eq.\ \eqref{eq:riemannian-limit}.

The last independent rank-three equation is found when we contract one of the last two indices of the full curvature tensor with the normal vector. In this case both the derivation and the final expression are much longer. Indeed, to derive it is useful to split the affine connection as the sum of the Levi-Civita part plus the distortion, and to reassemble the tensor structures at the very end. The outcome of these lengthy manipulations is the \emph{third generalized Codazzi relation}
\begin{empheq}[box={\mymath[colback=white!30, sharp corners]}]{equation}\label{eq:third-generalized-codazzi}
	\begin{split}
		R^\rho{}_{\sigma\mu\nu} h_{\rho\alpha} h^\sigma{}_\beta h^\mu{}_\lambda n^\nu = & \, \left[ \frac{1}{2} \left( D_\alpha + \epsilon \left( \kappa_\alpha - \tfrac{1}{2} \theta_\alpha \right) \right) \left( \Xi_{\beta\lambda} + 2 \Psi_{(\beta\lambda)} - 2 \Phi_{(\beta\lambda)} \right) - \left( \alpha \leftrightarrow \beta \right) \right]  \\
		& \, - \frac{1}{2} \left( D_\lambda - \epsilon \left( \kappa_\lambda - \tfrac{1}{2} \theta_\lambda \right) \right) \left( \Xi_{\alpha\beta} + 2 \Psi_{(\alpha\beta)} - 2 \Phi_{[\alpha\beta]} \right) + \epsilon \left( \kappa_\beta \Phi_{\lambda\alpha} - \lambda_\alpha K_{\lambda\beta} \right) \\
		& \, + L^{(3)}{}^\rho{}_{\beta\lambda} \Phi_{\rho\alpha} - L^{(3)}{}_{\alpha\lambda\rho} \Phi_\beta{}^\rho - h^\rho{}_\alpha h^\nu{}_\beta h^\mu{}_\lambda \pounds_n L^{(3)}{}_{\rho\nu\mu}  \, .
	\end{split}
\end{empheq}
This expression is entirely new and cannot be compared to other results in the literature; however, we can check that it reproduces the Riemannian result in the appropriate limit. To this end we notice that only the symmetric part of $\Phi^{\mu\nu}$ survives and the vector $\theta_\mu$ vanishes in this limit. Therefore, the two terms proportional to the $\kappa_\mu$ vector in the first line are exactly counterbalanced by the last terms in the second line, so that the final expression is indeed that which is found in the Riemannian case.

At this point one may use the previous expressions to find some appropriate relations for the contractions of the full curvature tensor, which is the non-metric case give rise to two different Ricci tensors. However, this is beyond the scope of the present work, and the such results can be worked out straightforwardly by using the generalized Gauss-Codazzi-Ricci equations that we have developed in this section.

\subsection{The scalar curvature}\label{subsect:scalar-curvature-relation}

While in a non-metric theory there are two independent Ricci tensors, there is only one nontrivial scalar, which we shall call full scalar curvature. Since this expression is that which appears in the $3+1$ decomposition of the Palatini action, we now focus on employing the previous results to find its expression. In order to expand the Ricci scalar in terms of the intrinsic and extrinsic geometric quantities we commence by writing is as
\begin{equation}\label{eq:gauss-codazzi-R-step1}
	\begin{split}
		R = & \, \left( \epsilon \, n^\mu n^\nu + h^{\mu\nu} \right) R_{\mu\nu}\\
		= & \, \left( \epsilon \, n^\mu n^\nu + h^{\mu\nu} \right) \left( \epsilon \, n^\lambda n^\rho + h^{\lambda\rho} \right) R_{\lambda\mu\rho\nu} \\
		= & \, \epsilon \left[ n^\mu n^\nu h^{\lambda\rho} R_{\lambda\mu\rho\nu} + n^\lambda n^\rho h^{\mu\nu} R_{\lambda\mu\rho\nu} \right] + h^{\mu\nu} h^{\lambda\rho} R_{\lambda\mu\rho\nu} \, .
	\end{split}
\end{equation}
The first two terms in the last line are the two contractions of the Ricci relations Eq.\ \eqref{eq:ricci-relation-1} and \eqref{eq:ricci-relation-2}, while the last one is recognized to be stemming from the Gauss equation \eqref{eq:gauss-relation}. In manipulating the first two expressions we find two Lie derivatives, whose sum can be rewritten as
\begin{equation}
	h^\nu{}_\mu \pounds_n \Phi_\nu{}^\mu - h^{\mu\nu} \pounds_n K_{\mu\nu} = \nabla_\mu \left[ n^\mu \left( \Phi^\nu{}_\nu - K^\nu{}_\nu \right) \right] + 2 K^{\mu\nu} \Phi_{\mu\nu} + \Phi^\mu{}_\mu \left( \Phi^\nu{}_\nu - K^\nu{}_\nu \right) + \epsilon \, \alpha \left( \Phi^\mu{}_\mu - K^\mu{}_\mu \right) \, .
\end{equation}
Furthermore, in both Ricci relations we find the covariant derivatives of two extrinsic non-metricity vectors, and we are able to improve these expressions by singling out the boundary terms, yielding
\begin{equation}
	h^{\mu\nu} D_\mu \kappa_\nu - h^\mu{}_\nu D_\nu \lambda^\mu = \nabla_\mu \left( \kappa^\mu - \lambda^\mu \right) + \epsilon \, \kappa_\mu \left( \kappa^\mu - \lambda^\mu \right) + \tilde{Q}^{(3)}{}_\mu \kappa^\mu \, .
\end{equation}
Finally, by combining all the previous expressions we can write down the Gauss-Codazzi relation for the full scalar curvature, which reads
\begin{equation}\label{eq:gauss-codazzi-R-final}
	\begin{split}
		R = &  \, R^{(3)} + 2 \alpha \, \Phi^a{}_a - \epsilon \, \Phi^{ab} \Phi_{ba} + \epsilon \, \Phi^a{}_a \Phi^b{}_b + \lambda^a \theta_a + \epsilon \, \kappa_a \tilde{Q}^{(3)}{}^a \\
		& + \epsilon \, \nabla_\mu \left[ n^\mu \left( \Phi^a{}_a - K^a{}_a \right) + e^\mu{}_a \left( \kappa_a - \lambda_a \right) \right] \, .
	\end{split}
\end{equation}
We have presented this result in such a way to show that the Riemannian limit is manifestly consistent with known results \cite{Poisson:2009pwt}. Indeed, by using Eq.\ \eqref{eq:riemannian-limit} we readily observe that the only non-vanishing terms in the first line those quadratic in $\Phi^{\mu\nu}$, which in Riemannian limit becomes $\Phi^{ab}\rightarrow -\oK^{ab}$,	thus reproducing the results of \cite{Poisson:2009pwt}. The same holds for the boundary terms too, in which the two vectors $\kappa_a \rightarrow a_a$ and $\lambda_a \rightarrow - a_a$, and they sum up to give the correct numerical factor in front of the acceleration vector.

\subsection{A different field basis and non-metricity scalars}\label{subsect:new-set}

In the presence of non-metricity the definition of the extrinsic curvature Eq.\ \eqref{eq:extrinsic-curvature} remains unchanged, whereas its actual form changes in that a contribution linear in the non-metricity appears in the second line of Eq.\ \eqref{eq:extrinsic-curvature}. This suggests us to adapt our basis in the space of rank-$2$ tensors on $\Sigma_t$ which is more suitable for separating the Riemannian and non-metricity contributions, which is given by the following definitions
\begin{eqnarray}\label{eqs:new-set-1}
	S_{ab} + A_{ab} \equiv n^\nu e^\rho{}_a e^\mu{}_b Q_{\rho\mu\nu} = K_{ab} + \Phi_{ab} \, , \quad && \Sigma_{ab} \equiv n^\rho e^\mu{}_a e^\nu{}_b Q_{\rho\mu\nu} = \Xi_{ab} + 2 \Psi_{(ab)} \, ,
\end{eqnarray}
where $S_{[ab]}=0$ and $A_{(ab)}=0$. To express the old variables in terms of the new ones we split the covariant gradient of the normal vector and employ the completeness relations, thus obtaining
\begin{equation}
	\begin{split}
		\nabla_\mu n_\nu = & \, \alpha \, n_\mu n_\nu + \epsilon \, n_\mu \kappa_\nu + \tfrac{\epsilon}{2} \theta_\mu n_\nu + K_{\mu\nu} \\
		= & \, \onabla_\mu n_\nu - L^\rho{}_{\mu\nu} n_\rho \\
		= & \, \alpha \, n_\mu n_\nu + \epsilon \, n_\mu \left( a_\nu + \tfrac{1}{2} \theta_\nu \right) + \tfrac{\epsilon}{2} \theta_\mu n_\nu + \oK_{\mu\nu} + S_{\mu\nu} - \tfrac{1}{2} \Sigma_{\mu\nu} \, .
	\end{split}
\end{equation}
 Accordingly, the extrinsic curvature and the extrinsic non-metricity tensor $\Phi_{ab}$ can be written as
\begin{eqnarray}\label{eqs:new-set-2}
	K_{ab} = \oK_{ab} - \frac{1}{2} \Sigma_{ab} + S_{ab} \, , \qquad && \Phi_{ab} = A_{ab} + \frac{1}{2} \Sigma_{ab} - \oK_{ab} \, ,
\end{eqnarray}
while the vector $\kappa_a$ may be written as
\begin{equation}
	\begin{split}
		\kappa_a = & \, \mathring{\kappa}_a + \tilde{\kappa}_a \\
		= & \, a_a + \tfrac{1}{2} \theta_a \, .
	\end{split}
\end{equation}
Since from Eq.\ \eqref{eq:riemannian-limit} we see that $\mathring{\kappa}_a = a_a$, we deduce that the non-metric part of the vector is $\tilde{\kappa}_a = \tfrac{1}{2} \theta_a$. The last expression will not be necessary in the following, whereas the Eq.\ \eqref{eqs:new-set-2} shall be employed thoroughly. Even though this new set of variables does not have a clear geometric origin as the ``old" one, it can be more convenient for practical calculations. Indeed, such a choice of basis will be very handy in order to carry out the Hamiltonian analysis and to obtain the correct Riemannian limit of the momenta conjugated to the intrinsic metric.

Endowed with these new definitions, we can focus on working out some suitable expressions for the decomposition of the five dimension-two scalars that are quadratic in the non-metricity, which is achieved by repeatedly using Eq.\ \eqref{eq:completeness-non-metricity}. The results are given by
\begin{subequations}
	\begin{align}\label{eq:Q-squared-decomp}
		Q^{\rho\mu\nu} Q_{\rho\mu\nu} = & \, 4 \epsilon \, \alpha^2 + 2 \left( \kappa_a + \lambda_a \right) \left( \kappa^a + \lambda^a \right) + \theta^a \theta_a + 2 \epsilon A_{ab} A^{ab} \\\nonumber
		& + 2 \epsilon S^{ab} S_{ab}  + \epsilon \Sigma_{ab} \Sigma^{ab} + Q^{(3)}{}_{cab} Q^{(3)}{}^{cab}  \, ; \\\label{eq:Q-twisted-squared-decomp}
		Q^{\rho\mu\nu} Q_{\mu\rho\nu} = & \, 4 \epsilon \, \alpha^2 + \left( 2 \theta^a + \kappa^a + \lambda^a \right) \left( \kappa_a + \lambda_a \right) - \epsilon A_{ab} A^{ab} \\\nonumber
		& + \epsilon S^{ab} S_{ab}  + 2 \epsilon S_{ab} \Sigma^{ab} + Q^{(3)}{}_{cab} Q^{(3)}{}^{acb} \, ; \\\label{eq:tr23Q-squared-decomp}
		Q_\rho Q^\rho = & \,4 \epsilon \, \alpha^2 + 4 \alpha \Sigma^a{}_a + \epsilon \Sigma^a{}_a \Sigma^b{}_b + \theta^a \theta_a \\\nonumber
		&  + 2 \epsilon \, \theta^a Q^{(3)}{}_a + Q^{(3)}{}_a Q^{(3)}{}^a \, ; \\\label{eq:tr12Q-squared-decomp}
		Q_\rho \tilde{Q}^\rho = & \,4 \epsilon \, \alpha^2 + 2 \alpha \left( S^a{}_a + \Sigma^a{}_a \right) + \epsilon S^a{}_a \Sigma^b{}_b + \epsilon \tilde{Q}^{(3)}{}_a \theta^a \\\nonumber
		& + \epsilon Q^{(3)}{}_a \left( \lambda^a + \kappa^a \right) + Q^{(3)}{}^a \tilde{Q}^{(3)}{}_a \, ; \\\label{eq:tr23Q-tr12Q-decomp}
		\tilde{Q}_\rho \tilde{Q}^\rho = & \, 4 \epsilon \, \alpha^2 + 4 \alpha S^a{}_a + \epsilon S^a{}_a S^b{}_b + \left( \kappa_a + \lambda_a \right) \left( \kappa^a + \lambda^a \right) \\\nonumber
		& + \tilde{Q}^{(3)}{}_a \tilde{Q}^{(3)}{}^a + 2 \epsilon \tilde{Q}^{(3)}{}_a \left( \kappa^a + \lambda^a \right)  \, .
	\end{align}
\end{subequations}
The equations that we have derived in this subsection are valid in every dimension, and for any non-null hypersurface $\Sigma$. In the next section we shall employ these results to the physical case of a time-space splitting of the manifold, thereby paving the route to the construction of the gravitational Hamiltonian in non-metric theories.

\subsection{The teleparallel limit}\label{subsect:telep-limit}

In the preceding parts of this section we have focused on the derivation of the generalized Gauss-Codazzi relations in a non-metric geometry, whereas now we want to concentrate on what happens when we impose the teleparallel constraint. The relevant and compelling feature of the analysis that we are about to carry out is that we shall not need any further assumption but the vanishing of the full curvature tensor, that is to say, we will \emph{not} work in the so-called coincident gauge. 

In a symmetric teleparallel geometry the total curvature vanishes, i.e.,
\begin{equation}
	R^\rho{}_{\lambda\mu\nu} = 0 \, .
\end{equation}
Therefore, since the left-hand sides of all the Gauss-Codazzi relations is zero, the same must be true for the right-hand sides as well. Consequently, it turns out that this requirement implies highly nontrivial constraints on the extrinsic tensors of the non-metric geometry. Let us start by considering the first and second generalized Codazzi relations Eqs.\ \eqref{eq:first-generalized-codazzi-2} and \eqref{eq:second-generalized-codazzi}: since these expressions are linear in $K_{\mu\nu}$ and $\Phi_{\mu\nu}$, respectively, we solve the teleparallel constraint by setting
\begin{eqnarray}\label{eq:telep-cond-rank-two}
	K_{\mu\nu} = 0 \, , \qquad && \Phi_{\mu\nu} = 0 \, .
\end{eqnarray}
The immediate consequence of these conditions is found by looking at the Gauss relation Eq.\ \eqref{eq:gauss-relation}, which sets the three-dimensional curvature to zero
\begin{equation}\label{eq:telep-cond-rank-four}
	R^{(3)}{}^d{}_{cab} = 0 \, .
\end{equation}
Therefore, the induced geometry is teleparallel as the four-dimensional one. Furthermore, two equations Eq.\ \eqref{eq:telep-cond-rank-two} also imply that the first and second generalized Ricci relations Eq.\ \eqref{eq:ricci-relation-1} and \eqref{eq:ricci-relation-2} are now homogeneous of degree one in the two vectors $\kappa_\mu$ and $\lambda_\mu$, respectively. Consequently, the teleparallel condition tells us that
\begin{eqnarray}\label{eq:telep-cond-rank-one-1}
	\kappa_\mu = 0 \, , \qquad && \lambda_\mu = 0 \, .
\end{eqnarray}
The story gets more complicated once we take into account the rank-one relation Eq.\ \eqref{eq:rank-one-relation} and the third generalized Codazzi relation Eq.\ \eqref{eq:third-generalized-codazzi}. The reason is that the previous constraints do not yield expressions which are linear in any of the remaining tensors. For this reason, one is generally not allowed to conclude that any of the tensors involved in these equations must vanish. Conversely, the deduction is that the Lie derivatives along the normal vector that appear in these relations can be written in terms of other tensor expressions. Indeed, from the rank-one relation Eq.\ \eqref{eq:rank-one-relation} we find
\begin{equation}\label{eq:telep-cond-rank-one-2}
	\pounds_n \theta_\mu = D_\mu \alpha + \frac{\epsilon}{2} \alpha \, \theta_\mu \, ,
\end{equation}
which indicates that \emph{no extrinsic vector is dynamical in any symmetric teleparallel theory}. On the other hand, the third generalized Codazzi relation Eq.\ \eqref{eq:third-generalized-codazzi} yields
\begin{equation}\label{eq:telep-cond-rank-three}
	h^\rho{}_\alpha h^\nu{}_\beta h^\mu{}_\lambda \pounds_n L^{(3)}{}_{\rho\nu\mu} = \frac{1}{2} \left( D_\alpha - \frac{\epsilon}{2} \theta_\alpha \right)  \Sigma_{\beta\lambda} - \left( D_{(\beta} - \frac{\epsilon}{2} \theta_{(\beta} \right)  \Sigma_{\lambda)\alpha} \, . 
\end{equation}
In hindsight, the fact that both hands of the previous equation are consistently symmetric in $\beta\leftrightarrow\lambda$ corroborates the result of the generalized Codazzi Eq.\ \eqref{eq:third-generalized-codazzi}. Furthermore, the fact that both the completely symmetric and the hook-symmetric parts of the previous equations are non-trivial tells us that \emph{no algebraic components of the induced distortion are dynamical in any symmetric teleparallel geometry}.

\section{The variational problem in non-metric theories}\label{sect:variational}

In this section we focus on deriving the correct form of the action functional in theories with non-metricity and, in particular, in the symmetric teleparallel equivalent theory of gravity. To this end, we need to correctly address the presence of boundary terms, whose inclusion in the action permits to construct a well-posed theory.

\subsection{The Palatini case}\label{subsect:variation-palatini}

Let us start by generalizing the purely Riemannian result for the Einstein-Hilbert action using the Palatini formulation, i.e., by considering the affine-connection as an independent field variable. In absence of non-metricity and torsion this result is derived in \cite{Poisson:2009pwt}, in which the boundary term is the trace of the extrinsic curvature. We observe that for $Q_{\mu\nu\rho}=0$ we have $\onabla_\mu n^\mu = \oK^a{}_a$, thus the boundary term can be written as the covariant divergence of the orthogonal vector $n^\mu$. On the other hand, this equality does not hold anymore if we have non-metricity. However, the presence of the divergence of $n^\mu$ is dictated by the fact that it cancels out the non-zero contribution to the field equations on the boundary in the variation of $R$. Thus, the Palatini action functional is given by
\begin{equation}\label{eq:palatini-action}
	S[g,\Gamma]_P = \frac{1}{16 \pi} \int_{\cal V} R \,\sqrt{-g}\, d^4 x + \frac{1}{8 \pi} \oint_{\partial {\cal V}} \epsilon \left( g^{\mu\nu} \nabla_\mu n_\nu  + \nabla_\mu n^\mu - K_0 + \Phi_0 \right) |h|^{1/2} d^3 y \, .
\end{equation}
The boundary term is such that it reproduces the term in Eq.\ \eqref{eq:gauss-codazzi-R-final} with the opposite sign. Furthermore, the subtracted terms in the last part of this equation are the trace of the extrinsic curvature and extrinsic non-metricity tensor $\Phi_{ab}$ of the manifold $\partial {\cal V}$ embedded in flat-space, and they are added by hand to avoid the presence of infinities when consider asymptotically flat spacetimes are taken into account. 

The variation of the Einstein-Hilbert term yields
\begin{equation}\label{eq:variation-R}
	\begin{split}
		\delta \int_{\cal V} R \,\sqrt{-g}\, d^4 x = & \int_{\cal V} \left[ \left( R_{\mu\nu} - \frac{1}{2} g_{\mu\nu} R \right) \delta g^{\mu\nu} + \left( Q_\lambda{}^{\mu\nu} - \delta_\lambda{}^\nu Q_\rho{}^{\mu\rho} \right) \delta \Gamma^\lambda{}_{\mu\nu} \right] \sqrt{-g} \, d^4 x\\
		& - \oint_{\partial {\cal V}} \epsilon \, n^\rho h^{\mu\nu} \partial_\rho \delta g_{\mu\nu} \, |h|^{1/2} d^3 y \, .
	\end{split}
\end{equation}
Notice that only the variation of the metric appears in the boundary integral, and that only its derivative along the normal $n^\mu$ can provide a non-trivial contribution for it is the only one which does not vanish. Such a boundary term is exactly canceled by the variation of the second term on the right-hand side of Eq.\ \eqref{eq:palatini-action}. Therefore, the full variation of the Palatini action in presence of non-metricity yields
\begin{equation}
	\delta S[g,\Gamma] = \frac{1}{16 \pi} \int_{\cal V} \left[ \left( R_{\mu\nu} - \frac{1}{2} g_{\mu\nu} R \right) \delta g^{\mu\nu} + \left( Q_\lambda{}^{\mu\nu} - \delta_\lambda{}^\nu Q_\rho{}^{\mu\rho} \right) \delta \Gamma^\lambda{}_{\mu\nu} \right] \sqrt{-g} \, d^4 x \, ,
\end{equation}
Accordingly, the field equations are given by
\begin{eqnarray}
	R_{\mu\nu} - \frac{1}{2} R g_{\mu\nu} = 0 \, , && \quad Q_{\rho\mu\nu} = 0 \, ,
\end{eqnarray}
which is exactly the result found by Palatini in \cite{Palatini:1919ffw}.

\subsection{The teleparallel case and the lambda symmetry}\label{subsect:variation-telep-eq}

Now we turn to teleparallel theories and concentrate on the variational problem. A virtue of the local approach to MAGs that has put forward by Adak and his collaborators \cite{Adak:2005cd} is provided by the useful insights regarding the introduction of Lagrange multipliers in the action functional. Indeed, in this language if we are to constrain the total curvature we are lead to add the four form $R^A{}_B \wedge \kappa^B{}_A$ to the Lagrangian, where $R^A{}_B$ is the gauge curvature two-form and $\kappa^B{}_A$ is the Lagrange multiplier two-form. In this way, the teleparallel condition is obtained from the field equation of the two-form $\kappa^B{}_A$. Moreover, since the differential Bianchi identities are $D R^A{}_B=0$, the Lagrange multiplier possesses a redundancy which gives rise to the so-called \emph{lambda symmetry} \cite{Adak:2005cd}.

When dealing with antisymmetric teleparallel theories, i.e., those in which the torsion is the only nontrivial field strength, one is forced to describe the geometry through the co-frame and the spin-connection. On the other hand, the functional dependencies of symmetric teleparallel thoeries may be expressed in terms of the metric $g_{\mu\nu}$ and a symmetric holonomic-connection $\Gamma^\rho{}_{(\nu\mu)}$. This is because there is no algebraic condition which can be imposed on a torsionful holonomic connection to set the non-metricity to zero. Conversely, such a simplification is granted by the symmetric nature of the non-metric full connection $\Gamma^\rho{}_{\lambda\mu}$ in a torsionless theory.

Given these premises, we can now write a four-form which is analogous to that which appears in \cite{Adak:2005cd} to impose the vanishing of the full-curvature, the only difference being the holonomic rather than non-holonomic nature of the $GL(d)$-valued indices. We write such a four-form as
\begin{equation}\label{eq:telep-action-LM}
	\mathcal{L}_{LM} = R^\rho{}_\lambda \wedge \kappa_\rho{}^\lambda = R^\rho{}_{\lambda[\mu\nu]} \kappa_\rho{}^\lambda{}_{[\alpha\beta]} \,\, dx^\mu \wedge dx^\nu \wedge dx^\alpha \wedge dx^\beta  \, .
\end{equation}
In a general torsionless theory the differential Bianchi identity takes the standard form, see Eq.\ \eqref{eq:bianchi-2}. Due to the symmetric nature of the connection, we can introduce an external differential operator which acts on tensor-valued $p$-forms such that it action is that of the external differential  $d$ on the $p$-form indices and it is given by that of $\nabla$ on the tensors-valued ones. Using this notation the Bianchi identities can be written as
\begin{equation}
	\nabla_{[\rho} R^\lambda{}_{|\sigma|\mu\nu]} \equiv \left( \nabla R^\lambda{}_\sigma \right)_{\rho\mu\nu} = 0 \, .
\end{equation}
Using such a formalism we can introduce the lambda symmetry in holonomic form, which is generated by the tensor-valued one-form $\lambda^\rho{}_\nu = \lambda^\rho{}_{\nu\mu} dx^\mu$. This symmetry transformation acts on the Lagrange multiplier as
\begin{equation}\label{eq:lambda-symmetry}
	\delta_{LM} \kappa^\rho{}_\lambda = \nabla \lambda^\rho{}_\lambda \, .
\end{equation}
By employing the rules for integrating by parts in the differential form formalism summarized in the App.\ \ref{app:sect:stokes} we see that this transformation is indeed a symmetry of the action. Thus, we take the functional form of the symmetric teleparallel action to depend on the metric tensor $g_{\mu\nu}$, the symmetric connection $\Gamma^\rho{}_{(\nu\mu)}$ and the Lagrange multiplier two-form $\kappa_\rho{}^\lambda{}_{[\mu\nu]}$, i.e.,
\begin{equation}\label{eq:functional-telep-action}
	S = S\left[g_{\mu\nu},\Gamma^\rho{}_{(\mu\nu)},\kappa_\rho{}^\lambda{}_{[\mu\nu]}\right] = \int \mathcal{L} \, ,
\end{equation}
where the full Lagrangian is then written as
\begin{equation}\label{eq:telep-action}
	\mathcal{L} = d^4 x \sqrt{-g} \,\, \barQ + R^\rho{}_\lambda \wedge \kappa_\rho{}^\lambda  \, .
\end{equation}
Here $\barQ$ is the general invariant quadratic expression in the non-metricity
\begin{equation}\label{eq:bar-Q}
	\barQ = c_1 \, Q^{\rho\mu\nu} Q_{\rho\mu\nu} + c_2 \, Q^{\rho\mu\nu} Q_{\mu\rho\nu} + c_3 \, Q_\rho Q^\rho + c_4 \, Q_\rho \tilde{Q}^\rho + c_5 \, \tilde{Q}_\rho \tilde{Q}^\rho \, .
\end{equation}
Let us now take the action to derive the field equations. In order to do that we first need an expression for the variation of the non-metricity, which is given by
\begin{equation}\label{eq:var-non-metricity}
	\begin{split}
		\delta Q_{\rho\mu\nu} = & \nabla_\rho \delta g_{\mu\nu} - 2 g_{\lambda(\nu} \delta \Gamma^\lambda{}_{\mu)\rho} \\
		= & - 2 Q_{\rho\alpha(\mu} g_{\nu)\beta} \delta g^{\alpha\beta} - g_{\mu\alpha} g_{\nu\beta} \nabla_\rho \delta g^{\alpha\beta} - 2 g_{\lambda(\nu} \delta \Gamma^\lambda{}_{\mu)\rho} \, .
	\end{split}
\end{equation}
We observe that the covariant derivative of $\delta g^{\mu\nu}$ is present in the final expression, while no such term appears for the affine connection, and the same is also true for the variation of $\barQ$, see Eq. \eqref{eq:var-barQ} in App.\ \ref{app:sect:variations}. Then, using the rules for integration by parts summarized in the App.\ \ref{app:sect:stokes}, we can write the variation of the first term in the action Eq.\ \eqref{eq:telep-action} as
\begin{equation}\label{eq:var-bar-Q}
	\begin{split}
		\delta \sqrt{-g} \,\, \barQ = & \sqrt{-g} \left[ \barq_{\mu\nu} - \frac{1}{2} g_{\mu\nu} \barQ \right] \delta g^{\mu\nu} + \sqrt{-g} Y^\rho{}_{\mu\nu} \nabla_\rho \delta g^{\mu\nu}  + \sqrt{-g} X_\rho{}^{\mu\nu} \delta \Gamma^\rho{}_{(\mu\nu)} \\
		= & \sqrt{-g} \left[ \barq_{\mu\nu} - \frac{1}{2} g_{\mu\nu} \barQ - \nabla_\rho Y^\rho{}_{\mu\nu} - \frac{1}{2} Q_\rho Y^\rho{}_{\mu\nu} \right] \delta g^{\mu\nu} + \sqrt{-g} X_\rho{}^{\mu\nu} \delta \Gamma^\rho{}_{(\mu\nu)}  \, ,
	\end{split}
\end{equation}
where the specific form of the tensors on the right-hand side depends on the actual theory that one picks up and it is not important for the present discussion. Finally, the variation of the Lagrange multiplier sector yields
\begin{equation}\label{eq:var-lagrange-multipl}
	\begin{split}
		\delta \left( R^\rho{}_\lambda \wedge \kappa_\rho{}^\lambda \right) = & \left( \nabla_{[\mu} \left( \delta \Gamma^\rho{}_{|\lambda|\nu]} \right) \kappa_\rho{}^\lambda{}_{[\alpha\beta]} + R^\rho{}_{\lambda[\mu\nu]} \delta \kappa_\rho{}^\lambda{}_{[\alpha\beta]}  \right) \,\, dx^\mu \wedge \dots \wedge dx^\beta \\
		= & \left( \delta \Gamma^\rho{}_{\lambda[\mu} \nabla_{\nu]} \kappa_\rho{}^{\lambda}{}_{[\alpha\beta]} +  R^\rho{}_{\lambda[\mu\nu]} \delta \kappa_\rho{}^\lambda{}_{[\alpha\beta]}  \right) \,\, dx^\mu \wedge \dots \wedge dx^\beta \, .
	\end{split}
\end{equation}
There are now three field equations that we must consider which stem from the variations that we have just compute. The first one that we take into account is that of the Lagrange multiplier, which simply yields the teleparallel condition
\begin{equation}\label{eq:lagrange-multip-field-eqs}
	R^\rho{}_{\lambda[\mu\nu]} = 0 \, .
\end{equation}
The huge advantage of the geometric formulation of the Lagrange multiplier sector of the action is that it does not concur to the metric field equations. Indeed from Eq.\ \eqref{eq:var-bar-Q} we can find their expression, which is
\begin{equation}\label{eq:metric-field-eqs-telep}
	\barq_{\mu\nu} - \frac{1}{2} g_{\mu\nu} \barQ - \nabla_\rho Y^\rho{}_{\mu\nu} - \frac{1}{2} Q_\rho Y^\rho{}_{\mu\nu} = 0 \, .
\end{equation}
On the other hand, the Lagrange multiplier does enter the field equations of the affine connection. In this regard, it is useful to write $vol\equiv \sqrt{-g}d^4 x$, so that by combining Eq.s\ \eqref{eq:var-barQ} and \eqref{eq:var-lagrange-multipl} we are able to cast these field equations as
\begin{equation}\label{eq:connection-field-eqs-telep}
	X_\rho{}^{(\lambda\mu)} \, vol + dx^{(\mu} \wedge \nabla \kappa_\rho{}^{\lambda)} \, .
\end{equation}
The last equation should be understood as determining the Lagrange multiplier as a function of the hypermomentum of $\barQ$. However, not all the components of $\kappa$ are determined, since the previous equation is invariant under a lambda transformation. Indeed, by varying the last equation along a lambda-transformation we find
\begin{equation}
	\begin{split}
		dx^{(\mu} \wedge \nabla \delta \kappa_\rho{}^{\lambda)} = & \, dx^{(\mu} \wedge \nabla \, \nabla \, \lambda_\rho{}^{\lambda)} \\
		= & \, dx^{(\mu} \wedge \left( R^{\lambda)}{}_\alpha \wedge \lambda_\rho{}^\alpha -   R^\alpha{}_\rho \wedge \lambda_\alpha{}^{\lambda)} \right) = 0 \, ,
	\end{split}
\end{equation}
where we have used Eq.\ \eqref{eq:lagrange-multip-field-eqs} in the last step. In a broader perspective the metric field equations are the only ones which actually dictate the dynamics of the specific symmetric teleparallel theory, whereas the other two field equations set the curvature to zero and generate the constraints out of which one should solve for the Lagrange multiplier.

Let us now focus on the boundary term in the action functional, which was discarded in Eq.\ \eqref{eq:telep-action}. For the sake of brevity we consider only the type of boundary term which appears in STEGR, which is
\begin{equation}
	{\cal B} = \int_{\cal V} \sqrt{-g} \mathring{\nabla}_\mu \left( Q^\mu - \tilde{Q}^\mu \right) \, .
\end{equation}
The other combinations can be derived straightforwardly using the machinery developed in the previous section \ref{sect:foliations}. Using the rules for integration by parts derived in the App.\ \eqref{app:sect:stokes} we can rewrite this term as an actual integral over the boundary, i.e.,
\begin{equation}
	\int_{\cal V} \sqrt{-g} \, \mathring{\nabla}_\mu \left( Q^\mu - \tilde{Q}^\mu \right) = \epsilon \int_{ \partial{\cal V}} \sqrt{|h|} \, n_\mu \left( Q^\mu - \tilde{Q}^\mu \right) \, .
\end{equation}
Thence, employing Eq.\ \eqref{eq:variation-boundary-vector} to evaluate the variation of this term, we find
\begin{equation}
	\begin{split}
		\delta \int_{\cal V}  \sqrt{-g} \mathring{\nabla}_\mu \left( Q^\mu - \tilde{Q}^\mu \right) = &\, \int_{ \partial{\cal V}}  \epsilon \sqrt{|h|} n_\mu \left[ \nabla_\nu \delta g^{\nu\mu} - g_{\alpha\beta} \nabla^\mu \delta g^{\alpha\beta} + \left( Q^\mu{}_{\alpha\beta} + \delta^\mu{}_{(\alpha} Q_{\beta)} \right) \delta g^{\alpha\beta} \right.\\
		& \left. + 2 \left( \delta^{(\rho}{}_\alpha g^{\mu)\lambda} - g^{\mu\rho} \delta^\lambda{}_\alpha \right) \delta \Gamma^\alpha{}_{\lambda\rho} \right] \, .
	\end{split}
\end{equation}
All undifferentiated variations of the fields vanish when evaluated on the boundary, thus all terms but the first two drop out of the previous equation. On the other hand, in these nontrivial two terms we can trade covariant derivatives for partial ones, since only the latter survive on the boundary. Thereby, we obtain
\begin{equation}
	\begin{split}
		\delta \int_{\cal V}  \sqrt{-g} \mathring{\nabla}_\mu \left( Q^\mu - \tilde{Q}^\mu \right) = \, \int_{ \partial{\cal V}}  \epsilon \sqrt{|h|} n^\mu h^{\nu\lambda} \partial_\mu \delta g_{\nu\lambda} \, .
	\end{split}
\end{equation}
Notice that this expression is exactly minus that which we found by varying the curvature scalar in the preceding subsection, see Eq.\ \eqref{eq:variation-R}. Thus, the boundary terms of Riemannian geometry and symmetric teleparallel gravity are identical, in that they give rise to the same nontrivial term upon variation. We further observe that this boundary term can be counterbalanced by including the following expression into the action
\begin{equation}\label{eq:boundary-counterterm-stegr}
	S_B = - \frac{1}{16 \, \pi} \oint_{\partial {\cal V}} \sqrt{|h|} \, \epsilon \, n^\rho h^{\mu\nu} \left( \nabla_\mu g_{\rho\nu} - \nabla_\rho g_{\mu\nu} \right) \, .
\end{equation}
Clearly, when computing the variation of this integral only the normal derivative of the metric variation gives a non-vanishing result, yielding the expected expression. In spite of the aforementioned analogy between GR and STEGR, we must make a remark on the huge difference between the two of them when it comes to boundary terms. Indeed, while in GR one is forced to include a boundary term in order to obtain a well-posed variational problem \footnote{There is a well-known way to avoid the presence of boundary terms in GR, which is that of starting from the so-called $\Gamma-\Gamma$ action. However, the price that one has to pay is the loss of manifest covariance.} , no such issue arises in symmetric teleparallel theories. This is because in the latter case one never meets the normal derivatives acting on $\delta g^{\mu\nu}$ evaluated on the boundary after computing the variation of any $f(\barQ)$ action. Thus, the inclusion of boundary terms is not required in symmetric teleparallel theories.

\section{The Hamiltonian analysis}\label{sect:hamiltonian-analyses}

In this section, we perform the Hamiltonian analysis of some non-metricity based theories of gravity. To do so, we commence by introducing the appropriate definition of time derivative in generally covariant theories. In order to give an unambiguous and geometrical meaning to these temporal derivatives, one defines them to be the Lie derivatives along the temporal vector field $t^\mu$ Eq.\ \eqref{eq:def-time-vector}. The flow along this vector field generates the dynamical evolution of the system, and any tensor field $\phi$ living on the constant-time hypersurfaces $\Sigma_t$ evolves according to $\dot{\phi}=\pounds_t \phi$. Then, the temporal derivatives are evaluated by using the fact that the basis vectors $e^\mu{}_a$ are invariant under this flow, i.e.,
\begin{equation}
	\pounds_t e^\mu{}_a = 0 \, .
\end{equation}
Thus, our first task is to work out an equation for the temporal Lie derivative of the metric tensor. Making use of the symmetry of the connection we write
\begin{equation}\label{eq:tdot-g}
	\begin{split}
		\pounds_t g_{\mu\nu} = & \, t^\alpha \nabla_\alpha g_{\mu\nu} + g_{\mu\alpha} \nabla_\nu t^\alpha + g_{\nu\alpha} \nabla_\mu t^\alpha \\
		= & \, \nabla_\mu t_\nu + \nabla_\nu t_\mu + t^\alpha \left( Q_{\alpha\mu\nu} - 2 Q_{(\mu\nu)\alpha} \right)\\
		= & \, N \left( \nabla_\mu n_\nu + \nabla_\nu n_\mu \right) + \nabla_\mu N_\nu + \nabla_\nu N_\mu + n_\mu \nabla_\nu N + n_\nu \nabla_\mu N \\
		& + N n^\alpha \left( Q_{\alpha\mu\nu} - 2 Q_{(\mu\nu) \alpha} \right) + N^\alpha \left( Q_{\alpha\mu\nu} - 2 Q_{(\mu\nu)\alpha} \right) \, .
	\end{split}
\end{equation}
To proceed further we need to find some suitable expressions of $\dot{N}$, $\dot{N}^a$ and $\dot{h}_{ab}$, which must enter the left-hand side of the previous equation. To this end, we start by exploiting the fact that $\pounds_t t^\mu = 0$ and $\pounds_t e^\mu{}_a=0$ to write
\begin{equation}\label{eq:tdot-n}
	\pounds_t n^\mu = - \frac{1}{N} n^\mu \pounds_t N - \frac{1}{N} e^\mu{}_a \pounds_t N^a \, .
\end{equation} 
Then, the desired time derivative are extracted from Eq.\ \eqref{eq:tdot-g} by contraction with the basis vectors $e^\mu{}_a$ and the normal vector. However, when we contract such an equation with $e^\mu{}_a n^\nu$ or $n^\mu n^\nu$ and we use Eq.\ \eqref{eq:tdot-n} we cannot extract the $\dot{N}^a$ or $\dot{N}$, but we only find trivial identities. This is because the lapse function $N$ and the shift vector $N_a$ are gauge-dependent objects that do not have an intrinsic physical meaning, but they only serve to parametrize the arbitrariness in picking up a specific realization of the temporal flow. Therefore, we only need to consider the contraction of Eq.\ \eqref{eq:tdot-g} with $e^\mu{}_a e^\nu{}_b$, which yields the desired expression for the time derivative of the intrinsic metric, i.e.,
\begin{equation}
	\dot{h}_{ab} = \, N \left( \Xi_{ab} + 2 \Psi_{(ab)} - 2 \Phi_{(ab)} \right) + D_a N_b + D_b N_a + N^c \left( Q^{(3)}{}_{cab} - 2 Q^{(3)}{}_{(ab)c} \right) \, .\
\end{equation}
In order to invert the previous equation to find the extrinsic curvature and non-metricity in terms of the temporal derivative of the intrinsic metric we need to use the set of variables defined in Subsect.\ \ref{subsect:new-set}. Indeed, by using Eqs.\ \eqref{eqs:new-set-1} and \eqref{eqs:new-set-2} we obtain
\begin{equation}\label{eq:tdot-h}
	\begin{split}
		\dot{h}_{ab} = & \,  2 \left[ N \oK_{ab} + D_{(a} N_{b)} + N_c L^{(3)}{}^c{}_{ab} \right] \\
		= & \, 2 \left( N \oK_{ab} + \oD_{(a} N_{b)} \right)  \, .
	\end{split}
\end{equation}
Then, since $\dot{h}_{ab}$ can be written in terms of the Riemannian extrinsic curvature, we can calculate the momenta conjugated to $h_{ab}$ by computing the partial derivatives of the Lagrange density using the chain rule as
\begin{equation}\label{eq:partial-doth-chain-rule}
	\frac{\partial {\cal L}}{\partial \dot{h}_{ab}} = \frac{\partial \oK_{cd}}{\partial \dot{h}_{ab}} \frac{\partial {\cal L}}{\partial \oK_{cd}} = \frac{1}{2N} \, \frac{\partial {\cal L}}{\partial \oK_{ab}} \, .
\end{equation}
On the other hand, we shall need a different kind of relation in order to deal with teleparallel theories. Indeed, using the fact that in a teleparallel geometry $K_{ab}=0$ and $\Phi_{ab}=0$ (see Eq.\ \eqref{eq:telep-cond-rank-two}), we can trade the Riemannian extrinsic curvature for the extrinsic non-metricity tensor $\Sigma_{ab}$ as
\begin{equation}
	\oK_{ab} \overset{telep.}{=} \frac{1}{2} \Sigma_{ab} \, .
\end{equation}
Accordingly, the teleparallel expression fo the temporal derivative of the intrinsic metric $h_{ab}$ is
\begin{equation}\label{eq:tdot-h-telep}
	\dot{h}_{ab} = \,  N \Sigma_{ab} + 2 \oD_{(a} N_{b)} \, ,
\end{equation}
and the partial derivative with respect to it reads
\begin{equation}\label{eq:partial-doth-chain-rule-telep}
	\frac{\partial {\cal L}}{\partial \dot{h}_{ab}} = \frac{\partial \Sigma_{cd}}{\partial \dot{h}_{ab}} \frac{\partial {\cal L}}{\partial \Sigma_{cd}} = \frac{1}{N} \, \frac{\partial {\cal L}}{\partial \Sigma_{ab}} \, .
\end{equation}

At this point we must warn the reader regarding the presence of a quite important ambiguity that arises in such a derivation, which points out the correct manner to take the teleparallel limit in the present formalism. As we have seen above in Subsect.\ \ref{subsect:telep-limit}, there are some relations among the extrinsic tensors which can be employed in the teleparallel limit of symmetric metric-affine theories. Thus, one might be tempted to exploit such a correspondence already at the level of the completeness relations for the covariant gradient of the normal vector and the non-metricity. This would yield the two following equations
\begin{subequations}\label{eq:tentative-completeness-relations}
	\begin{align}
		\nabla_\mu n_\nu = & \, \alpha \, n_\mu n_\nu + \frac{1}{2} \epsilon \, \theta_\mu n_\nu - \frac{1}{2} \Sigma_{\mu\nu} \, ,\\
		Q_{\rho\mu\nu} = & \, 2 \epsilon \, \alpha \, n_\rho n_\mu n_\nu + n_\mu n_\nu \theta_\rho + \epsilon n_\rho \Sigma_{\mu\nu} + Q^{(3)}{}_{\rho\mu\nu} \, .
	\end{align}
\end{subequations}
However, if we trust these relations and employ them in Eq.\ \eqref{eq:tdot-g} we find that
\begin{equation}
	N \left[ \left( \nabla_\mu n_\nu + \nabla_\nu n_\mu \right) + n^\alpha \left( Q_{\alpha\mu\nu} - 2 Q_{(\mu\nu) \alpha} \right) \right] = - \epsilon \, N n_{(\mu} \theta_{\nu)} \, ,
\end{equation}
which yields a vanishing contribution to $\dot{h}_{ab}$ when contracted twice with the spacelike basis of vector fields $e^\mu{}_a$. Clearly, this is both unacceptable and in sharp contrast with the preceding derivation. Accordingly, we deduce that using the completeness relations Eq.\ \eqref{eq:tentative-completeness-relations} at the \emph{before} expressing tensor expressions in terms of the extrinsic tensors or doing the two things in reverse order does not commute. Therefore, we conclude that the tentative relations put forward in Eq.\ \eqref{eq:tentative-completeness-relations} cannot be safely used, and that the teleparallel limit of the extrinsic tensors can only be taken at the very end of any calculation.

\subsection{Symmetric Palatini Hamiltonian}\label{subsect:palatini-hamiltonian}

As a warming up exercise we want to construct the Hamiltonian of Palatini gravity, and the starting point is to rewrite the bulk and boundary parts of the Ricci scalar Eq.\ \eqref{eq:gauss-codazzi-R-final} in a more suitable fashion. To this end, we employ the novel basis introduces in Subsect.\ \ref{subsect:new-set}, which yields
\begin{equation}
	\begin{split}
		R \overset{bulk}{=} & \, R^{(3)} + 2 \alpha \, \Phi^a{}_a  - \epsilon \, \Phi^{ab} \Phi_{ba} + \epsilon \, \Phi^a{}_a \Phi^b{}_b + \lambda^a \theta_a + \epsilon \, \kappa_a \tilde{Q}^{(3)}{}^a \\
		= & \, R^{(3)} + \alpha \left( \Sigma^a{}_a - 2 \oK^a{}_a \right) + \epsilon \left[ - \oK^{ab} \oK_{ab} + \oK^a{}_a \oK^b{}_b + \oK^{ab} \Sigma_{ab} - \oK^a{}_a \Sigma^b{}_b \right.\\
		& \left. + A^{ab} A_{ab} - \frac{1}{4} \Sigma^{ab} \Sigma_{ab} + \frac{1}{4} \Sigma^a{}_a \Sigma^b{}_b \right] + \lambda^a \theta_a + \epsilon \, \kappa_a \tilde{Q}^{(3)}{}^a \, .
	\end{split}
\end{equation}
On the other hand, by employing the same variables the boundary terms take the following form
\begin{equation}
	\begin{split}
		R \overset{boundary}{=} & \epsilon \, \nabla_\mu \left[ n^\mu \left( \Phi^a{}_a - K^a{}_a \right) + e^\mu{}_a \left( \kappa_a - \lambda_a \right) \right] \\
		= & \epsilon \, \nabla_\mu \left[ - 2 n^\mu \oK^a{}_a + e^\mu{}_a \left( \kappa_a - \lambda_a \right) \right] \, .
	\end{split}
\end{equation}
Therefore, the action functional of symmetric Palatini gravity is written as
\begin{equation}\label{eq:action-palatini-1}
	\begin{split}
		S_P[g,\Gamma] = & \frac{1}{16 \pi} \int_{t_1}^{t_2} dt  \left\{ \int_{\Sigma_t} N \sqrt{h} \, d^3 y \left[ R^{(3)} + \alpha \left( \Sigma^a{}_a - 2 \oK^a{}_a \right) + \oK^{ab} \oK_{ab} - \oK^a{}_a \oK^b{}_b - \oK^{ab} \Sigma_{ab}  \right.\right.\\
		& \left.\left. \qquad\qquad\qquad\qquad\qquad\quad - A^{ab} A_{ab} + \oK^a{}_a \Sigma^b{}_b + \frac{1}{4} \Sigma^{ab} \Sigma_{ab} - \frac{1}{4} \Sigma^a{}_a \Sigma^b{}_b + \lambda^a \theta_a + \epsilon \, \kappa_a \tilde{Q}^{(3)}{}^a  \right] \right.\\
		& \left. \qquad\qquad\;\;\;\;\; + \int_{S_t} N \sqrt{\sigma} \, d^2 \theta \left[ 2 \mathring{k}^A{}_A - r^a \left( \kappa_a - \lambda_a \right) \right] \right\} \, ,
	\end{split}
\end{equation}
where $\mathring{k}^A{}_A$ is the Riemannian extrinsic curvature of the two-dimensional spacelike surface at constant $r$. We can now compute the momentum conjugated to $\dot{h}_{ab}$ using the chain rule Eq. \eqref{eq:partial-doth-chain-rule}, obtaining
\begin{equation}
	16 \pi p^{ab} \equiv 16 \pi \frac{\partial {\cal L}}{\partial \dot{h}_{ab}} = \sqrt{h} \left[ \oK^{ab} - \oK^c{}_c h^{ab}  - \frac{1}{2} \Sigma^{ab} - \alpha h^{ab} + \frac{1}{2} \Sigma^c{}_c h^{ab} \right] \, ,
\end{equation}
whence the correct Riemannian limit is manifestly obtained. These relation must be inverted in order to find the velocity $\dot{h}_{ab}$ in terms of the momentum $p_{ab}$. The desired expression is
\begin{equation}\label{eq:ok-to-p}
	\oK^{ab} = \frac{16 \, \pi}{\sqrt{h}} \left( p^{ab} - \frac{1}{d-2} p^c{}_c h^{ab} \right) + \frac{1}{2} \Sigma^{ab} + \frac{1}{d-2} \alpha \, h^{ab} \, .
\end{equation}
Then, we define the canonical Hamiltonian density as
\begin{equation}
	\begin{split}
		{\cal H}_P \equiv & \, \left( p^{ab} \dot{h}_{ab} - \sqrt{-g} {\cal L_P} \right) \, ,
	\end{split}
\end{equation}
By defining $P=p^a{}_a$ the first term of the previous equation reads
\begin{equation}
	\begin{split}
		p^{ab} \dot{h}_{ab} = N \left[ \frac{32\pi }{\sqrt{h}} \left( p^{ab} p_{ab} - \frac{1}{d-2} P^2 \right) + p^{ab} \Sigma_{ab} + \frac{1}{d-2} \alpha\, P \right] + 2 p^{ab} \oD_a N_b \, .
	\end{split}
\end{equation}
Moreover, using Eq.\ \eqref{eq:ok-to-p} we find that minus the Lagrangian gives
\begin{equation}
	\begin{split}
		-\mathcal{L} = & N \left[ - \frac{16 \pi}{\sqrt{h}} \left( p_{ab} p^{ab} + \frac{1}{d-2} P^2 \right) -  \frac{4 }{d-2} \alpha P + \frac{1}{16\pi} \left( A_{ab} A^{ab}  - R^{(3)} \right.\right.\\
		& \qquad\left.\left.  + \frac{3 (d-1)}{d-2} \alpha^2  - \lambda^{a} \theta_{a} + Q^{(3)}{}^{b}{}_{ab} \kappa^{a} \right) \right] \, .
	\end{split}
\end{equation}
Thus, combining the previous expressions we finally obtain the $3+1$ formulation of the canonical Hamiltonian density of symmetric Palatini gravity
\begin{equation}
	\begin{split}
		\mathcal{H}_P = & \, N \left[ \frac{16 \pi}{\sqrt{h}} \left( p_{ab} p^{ab} -   \frac{1}{d-2} P^2 \right) + p^{ab} \Sigma_{ab} - \frac{2}{d-2} \alpha \, P  \right.\\
		& \left. + \frac{\sqrt{h}}{16\pi} \left( A_{ab} A^{ab}  + \frac{3 (d-1)}{d-2} \alpha^2  - R^{(3)} -  \lambda^{a} \theta_{a}   + \tilde{Q}^{(3)}_a \kappa^{a} \right) \right] + 2 \, p^{ab} \oD_a N_b \, .
	\end{split}
\end{equation}
The Hamiltonian is then obtained by integrating the previous equation over a spacelike hypersurface $\Sigma_t$. Due to the functional dependencies of this Hamiltonian, we can write its general variation as
\begin{equation}
	H_P = \int_{\Sigma_t} \left( \mathcal{P}^{ab} \delta h_{ab} + \mathcal{H}_{ab} \delta p^{ab} - \mathcal{C} \delta N - 2 \mathcal{C}_a \delta N^a \right) d^3 y \, .
\end{equation}
Furthermore, we can define $\hat{p}^{ab}\equiv16\pi p^{ab}$, so that the momentum variation of $H_P$ can be written in a concise way as
\begin{equation}
	\mathcal{H}_{ab} = \frac{2N}{\sqrt{h}} \left( \hat{p}_{ab} - \frac{1}{d-2} \hat{P} h_{ab} \right) + N \left( \Sigma_{ab} - \frac{2}{d-2} \alpha \, h_{ab} \right) + 2 \oD_{(a} N_{b)} \, .
\end{equation}
The other variations of the canonical Hamiltonian are found in a similar fashion, and they yield a dynamical system whose description requires the introduction of primary constraints in order to display a well-defined Legendre transform connecting the Lagriangian and Hamiltonian descriptions. However, such a construction in the present case lies beyond the scope of this work, and it will not be carried out here. The only observation that we make is that none of the intrinsic or extrinsic tensors of non-metric origin describes the temporal derivative of an intrinsic field; therefore, all these new field variables are to be interpreted as coordinates in the Hamiltonian formalism. Furthermore, all these new variables demand the introduction of an equal number of primary constraints in order to define their conjugated momenta. Then, as we will see in the next Subsection, one only has to check that these new primary constraints are first-class to prove that this theory has the same number of propagating d.o.f.\ of General Relativity.

\subsection{Symmetric teleparallel theories}\label{subsect:stegr-hamiltonian}

Let us now focus on the symmetric teleparallel theories, and, more specifically, to the GR-equivalent one. In writing the action we shall omit the teleparallel enforcing Lagrange multiplier, which is not needed once we express the right-hand side in terms of nontrivial teleparallel extrinsic and intrinsic fields only. Thus, the action which comprises also the boundary term Eq.\ \eqref{eq:boundary-counterterm-stegr}, which reads as follows after employing the completeness relations Eqs.\ \eqref{eq:completeness-non-metricity}, \eqref{eq:completeness-first-trace} and \eqref{eq:completeness-second-trace}  and choosing $L^{(3)}{}^c{}_{ab}$ as the independent intrinsic rank-three tensor
\begin{equation}\label{eq:symm-telep-action-1}
	\begin{split}
		S_{STEGR} = & - \frac{1}{16\pi} \int_{\cal V} \sqrt{-g} d^4x \, \left( \Q + \onabla_\mu \left( Q^\mu - \tilde{Q}^\mu \right) \right) + \frac{\epsilon}{16\pi} \oint_{\partial {\cal V}} \sqrt{|h|} d^3 y \, \Sigma \\
		= & - \frac{1}{16\pi} \int_{\cal V} \sqrt{-g} d^4x \, \mathbb{Q} - \frac{\epsilon}{16\pi} \oint_{\partial {\cal V}} \sqrt{|h|} d^3y \left[ n_\mu \left( Q^\mu - \tilde{Q}^\mu \right) - \Sigma \right] \\
		= & - \frac{1}{16\pi} \int_{t_1}^{t_2} dt \left\{ \int_{\Sigma_t} \sqrt{h} d^3y \, \mathbb{Q} \, + \, \int_{S_t} \sqrt{\sigma} d^2\theta N \left[ r_a \left( Q^{(3)}{}^a - \tilde{Q}^{(3)}{}^a - \theta^a \right) - \mathscr{S} \right]  \right\} \\
		= & \, \frac{1}{16 \pi} \int_{t_1}^{t_2} dt  \left\{ \int_{\Sigma_t} N \sqrt{h} \, d^3 y \left[ - \mathbb{Q}^{(3)} - \oD_a \left( Q^{(3)}{}^a - \tilde{Q}^{(3)}{}^a  \right) + \frac{1}{4} \Sigma_{ab} \Sigma^{ab} - \frac{1}{4} \Sigma^2 + \left( \tilde{L}^{(3)}{}_a - L^{(3)}{}_a \right) \theta^a  \right] \right.\\
		& \left.\qquad\qquad\quad + \int_{S_t} N \sqrt{\sigma} d^2\theta \left( \mathscr{S} + r_a \theta^a \right) \right\} \, .
	\end{split}
\end{equation}
Here, we have integrated by parts the linear combination of induced non-metricity vectors, highlighting the fact that it combines in the correct way to the induced non-metricity scalar. However, we still need to inspect carefully the boundary term. First of all, we can rearrange the boundary term using the fact that
\begin{equation}
	\begin{split}
		g_{\mu\nu} = r_\mu r_\nu + \gamma_{\mu\nu} = r_\mu r_\nu - n_\mu n_\nu + \sigma_{\mu\nu} \, ,
	\end{split}
\end{equation}
where $\sigma_{\mu\nu}$ is the two-dimensional induced metric on $S_t$. In this way the expression on the boundary can be written as
\begin{equation}
	\begin{split}
		\mathscr{S} + r^a \theta_a = & \, r^\rho \gamma^{\mu\nu} Q_{\rho\mu\nu} + r^\rho n^\mu n^\nu Q_{\rho\mu\nu}\\
		= & \, r^\rho \sigma^{\mu\nu} Q_{\rho\mu\nu} \\
		= & \, \varsigma^{AB} \sigma_{AB} \, ,
	\end{split}
\end{equation}
where $\varsigma^{AB}$ is the extrinsic non-metricity tensor of $S_t$. Furthermore, we employ Eqs.\ \eqref{eq:completeness-first-trace} and \eqref{eq:completeness-second-trace} from three to two dimensions to write
\begin{equation}
	Q^{(3)}{}_a - \tilde{Q}^{(3)}{}_a = e_a{}^A \left( Q^{(2)}{}_A - \tilde{Q}^{(2)}{}_A - \theta_A \right) + r_a \varsigma^A{}_A \, . 
\end{equation}
When we use this expression in Eq.\ \eqref{eq:symm-telep-action-1} we see that the extrinsic non-metricity tensor scalar of the two-dimensional cancels out, so that the integrand of the boundary term in the action vanishes identically. Thus, the complete action functional reads
\begin{empheq}[box={\mymath[colback=white!30, sharp corners]}]{equation}\label{eq:teleparallel-action-final-form}
		S_{STEGR} = \frac{1}{16 \pi} \int_{t_1}^{t_2} dt  \left\{ \int_{\Sigma_t} N \sqrt{h} \, d^3 y \left[ - \mathbb{Q}^{(3)} + \frac{1}{4} \Sigma_{ab} \Sigma^{ab} - \frac{1}{4} \Sigma^2 + \left( \tilde{L}^{(3)}{}_a - L^{(3)}{}_a \right) \theta^a \right] \right\} \, .
\end{empheq}
This is the well-posed action of the symmetric teleparallel equivalent of GR after the three-plus-one decomposition which will be used below.

We start our Hamiltonian analysis by computing the momentum, which is done using the chain rule Eq.\ \eqref{eq:partial-doth-chain-rule-telep} in Eq. \eqref{eq:teleparallel-action-final-form} and yields
\begin{equation}
	p^{ab} = \frac{\sqrt{h}}{32\pi} \left( \Sigma^{ab} - h^{ab} \Sigma \right) \, .
\end{equation}
Consequently, the extrinsic non-metricity is written in terms of the momentum as
\begin{equation}\label{eq:sigma-to-p}
	\Sigma^{ab} = \frac{32\pi}{\sqrt{h}} \left( p^{ab} - \frac{1}{d-2} P h^{ab} \right) \, ,
\end{equation}
where we have defined a shorthand for the trace of the momentum as $P=p^a{}_a$. Furthermore, even though our main concern is that of a four-dimensional theory, we keep $d$ as a unspecified integer for the sake of generality. Then, the canonical Hamiltonian is defined in the usual way, i.e.,
\begin{equation}
	\begin{split}
	\mathcal{H}_{ST} = & \, p^{ab} \dot{h}_{ab} - \mathcal{L}_{ST} \\
	= & \, p^{ab} \left( N \Sigma_{ab} + 2 \oD_{(a} N_{b)} \right) - \mathcal{L}_{ST} \, .
	\end{split}
\end{equation}
By combining Eq.\ \eqref{eq:sigma-to-p} and the previous one we can write the first term in the Hamiltonian density as
\begin{equation}
	p^{ab} \dot{h}_{ab} =  \frac{32\pi N}{\sqrt{h}} \left( p^{ab} p_{ab} - \frac{1}{d-2} P^2 \right) + 2 p^{ab} \oD_{(a} N_{b)} \, .
\end{equation}
Finally, we employ the completeness relations Eqs.\ \eqref{eq:completeness-non-metricity}, \eqref{eq:completeness-first-trace} and \eqref{eq:completeness-second-trace}, which gives the symmetric teleparallel equivalent canonical Hamiltonian 
\begin{equation}
	\begin{split}
		\mathcal{H}_{ST} = \frac{16 \pi \, N}{\sqrt{h}} \left( p^{ab} p_{ab} - \frac{1}{d-2} P^2 \right) + 2 p^{ab} \oD_{(a} N_{b)} + \frac{\sqrt{h}\, N}{16\pi} \left[ \Q^{(3)} + \left( L_a - \tilde{L}_a \right) \theta^a \right] \, .
	\end{split}
\end{equation}
Here we have opted for simplifying the notation, referring to the intrinsic distortion (and its traces) simply as $L^c{}_{ab}$, as we shall do till the end of this analysis. The previous Hamiltonian is a functional of the momentum $p^{ab}$ and the coordinates $h_{ab}$, $N^a$, $N$, $\theta_a$ and $L^c{}_{ab}$, therefore its variation can be written as
\begin{equation}
	\delta H_P = \int_{\Sigma_t} d^3 y \left( \mathcal{P}^{ab} \delta h_{ab} + \mathcal{H}_{ab} \delta p^{ab} + \mathcal{C} \delta N - 2 \mathcal{C}_a \delta N^a + \Theta^a \delta \theta_a + \Lambda_c{}^{ab} \delta L^c{}_{ab} \right)  \, .
\end{equation}
The last four terms the yield constraints, since the momenta conjugated to these variables do not appear in the canonical Hamiltonian. In particular, the last one is them reads
\begin{equation}
	\Lambda_c{}^{ab} = \frac{\sqrt{h}\, N}{16\pi} \left[ \delta^{(a}{}_c \left( L^{b)} - \frac{1}{2} \theta^{b)} \right) + h^{ab} \left( \tilde{L}_c + \frac{1}{2} \theta_c \right) - 2 L^{(ab)}{}_c \right] \, .
\end{equation}
On the other hand, the constraint of the extrinsic vector non-metricity is given by the following vector-density
\begin{equation}
	\Theta^a = \frac{\sqrt{h}\, N}{16\pi} \left( L^a - \tilde{L}^a \right) \, .
\end{equation}
From these two equations we deduce that it is possible to express the extrinsic non-metricity vector and the intrinsic distortion in terms of the constraints $\Lambda_c{}^{ab}$ and $\Theta^a$. Indeed, this is can be done because the three independent vectors in terms of $\Theta^a$ and the two traces of $\Lambda_c{}^{ab}$ can be solved for in the previous equation, yielding
\begin{subequations}\label{eq:theta-L-to-constraints}
	\begin{align}
		\theta_a = & \frac{16\pi}{(d-2)\sqrt{h} N} \left[ \Lambda_a{}^c{}_c - 2 \Lambda_c{}^c{}_a - 2 \Theta_a \right] \, ,\\
		\tilde{L}_a = & \frac{16\pi}{(d-2)\sqrt{h} N} \left[ \frac{1}{2} \Lambda_a{}^c{}_c + \Lambda_c{}^c{}_a - \Theta_a \right] \, , \\
		L_a = & \frac{16\pi}{(d-2)\sqrt{h}N} \left[ \frac{1}{2} \Lambda_a{}^c{}_c + \Lambda_c{}^c{}_a + (2d-5) \Theta_a \right] \, .
	\end{align}
\end{subequations}
Notice that, we do not encounter difficulties in expressing these vectors as linear combinations of the primary constraints but for the notable exception of $d=2$, in which GR itself becomes a topological theory with no dynamical degrees of freedom. Some analogous equations can be written for the trace-free parts of the distortion too, in such a way that all the algebraically independent parts of this tensor can be written as a linear combination of the constraints.

Now turn to the constraints that are present also in GR, i.e., the Hamiltonian and momentum constraints. These are those that we find by varying the Hamiltonian with respect to the lapse function and shift vector. In particular, the expression of latter constraint is exactly the same as in Riemannian geometry, i.e.,
\begin{equation}
	\mathcal{C}_a = \sqrt{h} \oD_b p_a{}^b \, ,
\end{equation}
whereas the former yields
\begin{equation}
	\mathcal{C} = \frac{16\pi}{\sqrt{h}} \left( p^{ab} p_{ab} - \frac{1}{d-2} P^2 \right) + \frac{\sqrt{h}}{16\pi} \left[ L_a \tilde{L}^a + \frac{1}{2} \left( L_a - \tilde{L}_a \right) \theta^a - L_{cab} L^{acb} \right] \, .
\end{equation}
On the other hand, in contrast with the last result and analogously to the momentum constraint, we observe that by varying with respect to $p^{ab}$ we find the same expression of Riemannian geometry, that is
\begin{equation}
	\mathcal{H}_{ab} = \frac{32\pi}{\sqrt{h}} \left( p_{ab} - \frac{1}{d-2} h_{ab} P \right) + 2 \oD_{(a} N_{b)} \, .
\end{equation}
Lastly, the variation along $h_{ab}$ of the Hamiltonian gives rise to
\begin{equation}
	\begin{split}
		\mathcal{P}^{ab} = &  \,  \frac{16\pi N}{\sqrt{h}} \left[ 2 \left( p^a{}_c p^{cb} - \frac{1}{d-2} P p^{ab} \right) - \frac{1}{2} \left( p^{cd} p_{cd} - \frac{1}{d-2} P^2 \right) h^{ab} \right] + 2 p^{c(a} \oD_c N^{b)} - p^{ab} \oD_c N^c - N^c \oD_c p^{ab} \\
		& + \frac{\sqrt{h}\, N}{16\pi} \left[ L^{cd(a} L_{dc}{}^{b)} - \tilde{L}_c L^{cab} - L_c{}^{ab} \theta^c - \left( L^{(a} - \tilde{L}^{(a} \right) \theta^{b)} \right.\\
		& \left.\qquad\qquad + \frac{1}{2} \left( L_c \tilde{L}^c - L_{cde} L^{dce} + \left( L_c - \tilde{L}_c \right) \theta^c \right) h^{ab} \right] \, .
	\end{split}
\end{equation}
In the derivation process of the preceding equations we have discarded the boundary contributions that one finds when variating the Hamiltonian. However, these terms can be integrated by parts and evaluated on the boundary $S_t$ of the three-dimensional hypersurfaces $\Sigma_t$. This implies that, since in this way no normal derivatives of the fields variations appear, these boundary contributions do not affect the present analysis as long as the fields are kept fixed on the boundary, and we shall henceforth assume that this is indeed the case.

At this point it is convenient to cast the constraints and tensor-densities that we have just derived into a more convenient form, which highlights the simplification that occur in the Hamiltonian analysis. In particular, since the formal dependence on the momentum $p^{ab}$ is exactly the same that we find in GR, we can split all the constraints and variations of the Hamiltonian so that they are the sum of a Riemannian part plus some deviations. Thus, we have that the Hamiltonian and momentum constraints can be written as
\begin{eqnarray}
	\mathcal{C} = \mathring{\mathcal{C}} + \frac{\sqrt{h}}{16\pi} \, c(L,\theta) \, , && \mathcal{C}_a = \mathring{\mathcal{C}}_a \, ,
\end{eqnarray}
whereas the variations with respect to the intrinsic metric and its conjugated momentum read
\begin{eqnarray}
	\mathcal{P}^{ab} = \mathring{\mathcal{P}}^{ab} + \frac{\sqrt{h}}{16\pi} \Upsilon^{ab} (L,\theta) \, \, && \mathcal{H}_{ab} = \mathring{\mathcal{H}}_{ab} \, .
\end{eqnarray}
Here $c(L,\theta)$ and $\Upsilon^{ab} (L,\theta)$ parametrize the deviations from GR and depend algebraically on the intrinsic distortion and the extrinsic vector. Furthermore, the latter tensors are understood to be functions of the constraints as in Eq.\ \eqref{eq:theta-L-to-constraints}. The same kind of split can also be performed in the canonical Hamiltonian, for which we have
\begin{equation}
	H_{ST} = \mathring{H} + f(\Theta^a, \Lambda_c{}^{ab},N,h) \, .
\end{equation}
Notice that the $f$ function is homogeneous of second degree in the constraints $\Theta^a, \Lambda_c{}^{ab}$. This implies that the following equation holds
\begin{equation}
	\{H,\Omega\} \approx \{\mathring{H},\Omega\} \, ,
\end{equation}
where $\Omega$ is an arbitrary function of the canonical variables.

Due to the absence of temporal derivatives of the $\left( N,N^a,\theta_a,L^c{}_{ab} \right)$ in the Lagrangian, we cannot express the coordinates and velocities in terms of the phase space variables $(q^n,p_n)$. Therefore, we are forced to introduce the primary constraints, which allow to perform the Legendre transformation providing a consistent alterative Hamiltonian description of the same Lagrangian dynamics. Thus, we first consider the primary constraints
\begin{eqnarray}\label{eq:old-primary-constraints}
	\phi_0 \equiv \pi_0 = 0 \, , && \phi_1{}_a \equiv \pi_1{}_a = 0 \, ,
\end{eqnarray}
which are chosen to coincide with the canonical momenta conjugated to the lapse function $N$ and shift vector $N^a$, respectively. In an analogous way, we introduce the primary constraints which enforce the vanishing of the canonical momenta of the $\theta_a$ and $L^c{}_{ab}$ tensors, i.e.,
\begin{eqnarray}\label{eq:new-primary-constraints}
	\phi_\theta{}^a \equiv \pi_\theta{}^a = 0 \, , && \phi_L{}_c{}^{ab} \equiv \pi_L{}_c{}^{ab} = 0 \, .
\end{eqnarray}
The coordinates and momenta of the phase space variables satisfy the standard Poisson brackets
\begin{equation}\label{eq:standard-poisson-brackets}
	\{q^i,p_j\} = \delta^i{}_j \, , \quad \{q^i,q_j\} = 0 \, , \quad \{p_i,p_j\} = 0 \, ,
\end{equation}
where $\delta^i{}_j$ is a shorthand which stands for the product of a three-dimensional delta times the identity of the appropriate tensor bundle. Now we are ready to present the modified Hamiltonian, which is done by taking into account a set $u^m$ of Lagrange multipliers which enforce the constraints through their contribution to the variational principle. This yields:
\begin{equation}\label{eq:modified-hamiltonian-stegr}
	\begin{split}
		\tilde{H} = & \, H + u^m \phi_m \\
		= & \, H +u^0 \phi_0 + u^a \phi_1{}_a + v^a \pi_\theta{}^a + w^c{}_{ab} \pi_L{}_c{}^{ab} \, .
	\end{split}
\end{equation}
Since the primary constraints must be true at all times, one must have that their temporal derivatives vanish. This requirement can be imposed using the Poisson bracket as
\begin{equation}\label{eq:secondary-constraints-condition}
	\dot{u}_m = \{ u_m , H \} + u^{m'} \{ \phi_m , \phi_{m'} \} \, .
\end{equation}
Whenever this equation is not satisfied automatically \emph{after} having used the equations of motion we are dealing with some secondary constraints. Thus, let us analyze wether the new primary constraints that we were forced to introduce give rise to secondary ones. When it comes to the Poisson bracket of $\pi_\theta{}^{a}$ with the canonical Hamiltonian we have that it vanishes weakly (see Eq.\ \eqref{eq:weak-equality})
\begin{equation}
	\{ \pi_\theta{}^{a} , H \} = - \frac{\partial H}{\partial \theta_a} \approx 0 \, ,
\end{equation}
while the second term in Eq.\ \eqref{eq:secondary-constraints-condition} clearly vanishes due to Eq.\ \eqref{eq:standard-poisson-brackets}. In a completely analogous way we also have that
\begin{equation}
	\{ \pi_L{}_c{}^{ab} , H \} = - \frac{\partial H}{\partial L{}^c{}_{ab}} \approx 0 \, .
\end{equation}
Thus, both newly introduced primary constraints Eq.\ \eqref{eq:new-primary-constraints} do not give rise to secondary ones. Furthermore, since they commute with all the other $\phi_m$, they are also first-class constraints. Thus, the current description of the symmetric teleparallel equivalent theory differs from General Relativity only through the presence of the canonical variables $\left( \theta_a , L^c{}_{ab} \, ; \pi_\theta{}^a , \pi_L{}_c{}^{ab} \right)$, whose total number is given by
\begin{equation}
	\#  \,\,{\rm new} \,\,{\rm canonical} \,\, {\rm  variables} = \# \left( \theta_a , L^c{}_{ab} \, ; \pi_\theta{}^a , \pi_L{}_c{}^{ab} \right) = 6 + 36 = 42 \, .
\end{equation}
On the other hand, we have just seen that the new momentum variables give rise to first-class primary constraints only. Thus, the total number of new first-class constraints is
\begin{equation}
	\#  \,\,{\rm new} \,\,{\rm first}-{\rm class} \,\, {\rm  constraints} = \# \left( \pi_\theta{}^a , \pi_L{}_c{}^{ab} \right) = 3 + 18 = 21 \, .
\end{equation}
Therefore, the canonical variables $ \left( \theta_a , L^c{}_{ab} \, ; \pi_\theta{}^a , \pi_L{}_c{}^{ab} \right)$ do not generate any new degree of freedom, and the standard GR computation of the d.o.f.\ is carried over unambiguously due to the formal coincidence of the symmetric teleparallel equivalent canonical Hamiltonian with the GR one.

\section{Discussion and Conclusions}\label{sect:conclusions}

In this work, we have adopted the formalism of the $3+1$ decomposition following the footsteps by \cite{Poisson:2009pwt, Jackson, Kluson:2013hza} to study the properties of foliations in non-metric geometries. In this context, our first new contribution is the generalized Gauss-Codazzi relations in Subsect.\ \ref{subsect:generalized-gauss-codazzi}. These are more complicated than in the Riemannian case, since there are many more extrinsic tensors to play with, and the different symmetry properties of the full curvature tensor allow more independent combinations of them to appear on the right-hand sides of the Gauss-Codazzi relations. Furthermore, we have analyzed the consequences of imposing the teleparallel condition on the extrinsic and intrinsic geometric objects introduced before. The result of this analysis is that there are two extrinsic quantities which can give rise to dynamical d.o.f.\ in a symmetric teleparallel theories. These are given by a symmetric rank-two tensor, which plays the same role of the extrinsic curvature in Riemannian geometry, and by a scalar which identically decouples when the STEGR case is examined. Moreover, an extrinsic vector and the intrinsic distortion are nontrivial in these theories, though their temporal derivatives \emph{cannot} appear in the theory as a consequence of the Gauss-Codazzi relations.

In Sect.\ \ref{sect:variational}, we have outlined the details of the variational problems in Palatini theory and STEGR, paving particular attention to the presence of boundary terms and their role in yielding a well-defined variational problem. Besides, we have also proposed a novel holonomic-based formulation of the teleparallel constraint in the Lagrangian formalism, combining the language of differential forms used by \cite{Adak:2005cd} with the coordinate-based approach of \cite{BeltranJimenez:2019esp}. In turn, such a new formulation of the teleparallel condition has lead us to notice the presence of a lambda-symmetry affecting the Lagrange multiplier two-form, which was previously noted only in the anholonomic framework \cite{Adak:2005cd}.

In the subsequent section \ref{sect:hamiltonian-analyses}, we have exploited the previous results to build the Hamiltonian densities for Palatini gravity and STEGR. The first step in this direction has been the correct identification of the temporal derivative of the intrinsic metric, in which we have used the previously obtained teleparallel limit of the extrinsic tensors. In particular, along such a derivation, we have observed an ambiguity that arises in teleparallel theories: the final result seems to depend on the order in which the computations are performed. Thus, this formalism has pushed us to select a protocol in order to solve this ambiguity. We have argued that this problem only arises when the covariant derivative of the normal vector appears, and that it is not present when the tensor expressions depend only on the non-metricity. Indeed, this is exactly the case in which the ambiguity has arisen in Eq.\ \eqref{eq:tentative-completeness-relations}. Therefore, we have been brought to deduce that the teleparallel limit can be carefully taken into account only at the very end of the  $3+1$ decomposition, i.e., only when all the tensors are expressed using the suitable basis of three-dimensional-index tensors.

After having settled the paramount issue of the protocol to perform the teleparallel limit, we have turned to two specific Hamiltonian theories. In the first one  (Subsect.\ \ref{subsect:palatini-hamiltonian})  we have employed the previously obtained results for the $3+1$ decomposition of the scalar curvature to write down the Hamiltonian density. Even though we have not dealt explicitly with the subsequent steps of the Hamiltonian analysis, we have outlined the procedure which needs to be followed for formally proving that the number of propagating d.o.f.\ is the same of GR.

In the second application, we have studied in detail the STEGR case. This has been done by firstly writing down the Lagrangian with the boundary term induced equivalent to that of GR. However, in this case we have seen that the final expression of the boundary integrand vanishes identically, thus marking a huge difference with GR, in that adding this boundary term to $\mathbb{Q}$ does not alter at all neither the variational principle nor the final form of the $3+1$-decomposed action Eq.\ \eqref{eq:teleparallel-action-final-form}. In particular, this functional depends on the extrinsic non-metricity $\Sigma_{ab}$, which plays the role of the extrinsic curvature in GR for it provides a geometric expression of the time derivative of the induced metric $h_{ab}$. Along with this tensor, the action also depends on an extrinsic vector $\theta_a$ and the three-dimensional distortion tensor $L^{(3)}{}^c{}_{ab}$. However, the analysis of Subsect.\ \ref{subsect:telep-limit} shows that the Lie derivatives of these tensors along the normal vector $n^\mu$ are not independent expressions, and that they can be written in terms of spatial derivatives. Thus, none of these two tensors can display nontrivial momenta at the Hamiltonian level. Consequently, starting from this Lagrangian density we have derived the canonical Hamiltonian and we have computed its first functional variation. As we have already alluded to, the absence of momenta for $\theta_a$ and $L^{(3)}{}^c{}_{ab}$  forced us to introduce some Lagrange multipliers in the modified Hamiltonian, in such a way that the primary constraints $\pi_\theta{}^a$ and $\pi_L{}^c{}_{ab}$ vanish on shell. Moreover, the subsequent analysis has also proved that these constraints are of first-class, whereas we have noticed that the rest of the Hamiltonian is formally identical to the GR one. Therefore, we have adopted the standard computation of  dynamical d.o.f.\ in constrained Hamiltonian systems to conclude that $\pi_\theta{}^a$ and $\pi_L{}^c{}_{ab}$ do not concur in any way to this calculation, thus yielding the same number of d.o.f \  that is found in GR. Finally, we remark that the highly compelling feature of this analysis is that at no point we made any approximation or gauge fixing, thus the final result is a true physical prediction.

Along with the previous lines, we will now point out which may be, from our own viewpoint, the most intriguing and compelling generalizations of the present work. First of all, we argue that the methods and formalism that we have proposed in this paper can be applied also to other specific models of symmetric teleparallel gravity. Indeed, due to the fact that at no point we were required to select a gauge, this framework provides a completely general tool for studying the Hamiltonian structure and the number of propagating d.o.f.\ in any symmetric teleparallel theory. In this sense, a very interesting application is that of $f(\mathbb{Q})$ theories, which have been employed intensely in cosmological applications but whose thorough Hamiltonian analysis is still lacking.

On the other hand, the absence of a boundary term in the final expression of the STEGR action has given rise to a huge difference with respect to GR, which should be further investigated in future studies. As a matter of fact, when computing the energy (more generally, the conserved quantities) in an asymptotically flat spacetime one needs to subtract the flat-space result in order to obtain a finite quantity, and this Minkowski result is inherently present in the GR Hamiltonian from the very beginning due to the presence itself of a non-trivial boundary term \cite{Poisson:2009pwt}. Conversely, since no such a term appears in the STEGR action, one faces, in principle, a problem of paramount importance when computing the energy of, say, a static black hole at infinity.

Finally, the present analysis yields the necessary tools for studying the symmetric teleparallel equivalent versions of other Riemannian theories of gravity that have attracted attention over the past years. Of particular relevance in this sense are the Starobinsky theory of inflation \cite{Starobinsky:1980te} and the higher-derivative generalizations of GR due to Stelle \cite{Stelle:1976gc,Stelle:1977ry}. In particular, teleparallel analog of the former needs a thorough comprehension of the realization of Weyl invariance in this geometric setting and its connection to scale and conformal invariance in the flat-space limit. While the study of such models is commonly motivated by high-energy field-theoretical motivations such as the presence of a conformal anomaly \cite{Christensen:1977jc,Riegert:1984kt,Deser:1993yx} or the need for renormalizability \cite{Stelle:1976gc,Anselmi:2017lia,Anselmi:2017ygm}, we do believe that performing the Hamiltonian analysis of symmetric teleparallel theories, resembling these models, can benefit our understanding of teleparallel theories as a whole, and shed light on the possible high-energy description of gravitational phenomena using the teleparallel paradigm.

\section*{Acknowledgments}
SC acknowledges the {\it Istituto Nazionale di Fisica Nucleare} (INFN) Sez.\ di Napoli, {\it Iniziative Specifiche} QGSKY and MoonLight-2  and the {\it Istituto Nazionale di Alta Matematica} (INdAM), gruppo GNFM, for the support. This paper is based upon work from COST Action CA21136 -- Addressing observational tensions in cosmology with systematics and fundamental physics (CosmoVerse), supported by COST (European Cooperation in Science and Technology).

The work of DS was founded by a {\it Fondazione Angelo della Riccia} grant. DS also acknowledges the {\it Theoretisch-Physikalisches Institut} of the {\it Friedrich-Schiller-Universität Jena} for its hospitality and for its financial support. DS would also like to thank Dra\v{z}en Glavan for a fruitful discussion regarding the topic of this paper.


\appendix


\section{Integration by parts}\label{app:sect:stokes}

In this appendix we derive the rules for integration by parts in a non-metric geometry. In particular, we focus on bulk integrals which are written in both the standard holonomic way and in the differential-form formalism. Let $I,J$ run over some unspecified set of internal, holonomic or anholonomic indices. Then, the Riemannian result for the Stokes theorem reads
\begin{equation}
	\int_{\cal V} \sqrt{-g} w^I \onabla_\mu v^\mu{}_I = - \int_{\cal V} \sqrt{-g} v^\mu{}_I \onabla_\mu w^I + \int_{ \partial{\cal V}} \epsilon \, |h|^{1/2} n_\mu w^I v^\mu{}_I \, .
\end{equation}
Therefore, we can find the corresponding expression in a non-metric geometry by using the splitting of the affine connection Eq.\ \eqref{eq:gamma-split}. To this end we first write
\begin{equation}
	\nabla_\mu v^\mu{}_I = \onabla_\mu v^\mu{}_I + L^\nu{}_{\mu\nu} v^\mu{}_I \, ,
\end{equation}
whence, we straightforwardly obtain
\begin{equation}\label{eq:integration-by-parts}
	\int_{\cal V} \sqrt{-g} w^I \nabla_\mu v^\mu{}_I = - \int_{\cal V} \sqrt{-g} v^\mu{}_I \left(\nabla_\mu w^I - L^\nu{}_{\mu\nu} w^I \right) + \int_{ \partial{\cal V}} \epsilon \, |h|^{1/2} n_\mu w^I v^\mu{}_I \, .
\end{equation}
By further using $L^\nu{}_{\mu\nu}=-\tfrac{1}{2} Q_\mu$ we can rewrite the previous equation as
\begin{equation}\label{eq:integration-by-parts2}
	\int_{\cal V} \sqrt{-g} w^I \nabla_\mu v^\mu{}_I = - \int_{\cal V} \sqrt{-g} v^\mu{}_I \left(\nabla_\mu + \frac{1}{2} Q_\mu \right) w^I + \int_{ \partial{\cal V}} \epsilon \, |h|^{1/2} n_\mu w^I v^\mu{}_I \, .
\end{equation}

Let us now take into account the outcome of integration by parts in the tensor-valued $p$-form formalism. In particular, we consider the tensor-valued one-form $\Psi^\lambda{}_\rho = \Psi^\lambda{}_{\rho\mu} dx^\mu$ and two-form $\Phi^\rho{}_\lambda = \Phi^\rho{}_{\lambda[\mu\nu]} dx^\mu \wedge dx^\nu$. Then, up to boundary terms, we have that
\begin{equation}\label{eq:integration-by-parts-forms}
	\begin{split}
		\nabla \Psi^\lambda{}_\rho \wedge \Phi^\rho{}_\lambda = & \left( d \Psi^\lambda{}_\rho + \Gamma^\lambda{}_\sigma \wedge \Psi^\sigma{}_\rho - \Gamma^\sigma{}_\rho \wedge \Psi^\lambda{}_\sigma \right) \wedge \Phi^\rho{}_\lambda \\
		= & \left( \partial_\mu \Psi^\lambda{}_{\rho\nu} + \Gamma^\lambda{}_{\sigma\mu} \Psi^\sigma{}_{\rho\nu} - \Gamma^\sigma{}_{\rho\mu} \Psi^\lambda{}_{\sigma\nu} \right) \Phi^\rho{}_{\lambda\alpha\beta} \,\, dx^\mu \wedge dx^\nu \wedge dx^\alpha \wedge dx^\beta \\
		= &  \Psi^\lambda{}_{\rho\nu} \left( - \partial_\mu \Phi^\rho{}_{\lambda\alpha\beta} + \Gamma^\sigma{}_{\lambda\mu} \Phi^\rho{}_{\sigma\alpha\beta} - \Gamma^\rho{}_{\sigma\mu} \Phi^\sigma{}_{\lambda\alpha\beta} \right) dx^\mu \wedge dx^\nu \wedge dx^\alpha \wedge dx^\beta \\
		= & \Psi^\lambda{}_\rho \wedge \nabla \, \Phi^\rho{}_\lambda \, .
	\end{split}
\end{equation}
Notice that the plus sign in the last line depends on the fact that $\Psi$ is an odd differential form. Thus, from the previous equation we observe that the covariant extension $\nabla$ of the exterior differential operator $d$ and $d$ itself share the same rules for integration by parts. In particular, at no point the fact that the geometry is non-metric enters the previous equation. This contrasts with Eq.\ \eqref{eq:integration-by-parts2}, and it is due to the fact that there is no need to have a metric structure in order to define differential forms and a symmetric connection $\nabla$ which extends $d$. Furthermore, this result clearly holds true for tensor-valued $p$-forms of any kind.

\section{Variations}\label{app:sect:variations}

In this appendix we present the variations and $3+1$ decompositions of all the scalars of mass dimension-two quadratic in the non-metricity. Though we shall not specifically deal with a teleparallel theory, all of the results that follow can be generalized to that case in a simple manner.

For the variation of the bulk terms we have that
\begin{subequations}\label{eq:var-Q-single-pieces}
	\begin{align}
		\delta \left( Q^{\rho\mu\nu} Q_{\rho\mu\nu} \right) = & \left( Q_{\alpha\mu\nu} Q_\beta{}^{\mu\nu} - 2 Q^{\mu\nu}{}_\alpha Q_{\mu\nu\beta} \right) \delta g^{\alpha\beta} - 2 Q^\rho{}_{\alpha\beta} \nabla_\rho \delta g^{\alpha\beta} - 4 Q^{(\mu\nu)}{}_\lambda \delta \Gamma^\lambda{}_{\nu\mu} \, , \\
		\delta \left( Q^{\rho\mu\nu} Q_{\mu\rho\nu} \right) = & - Q^{\mu\nu}{}_{(\alpha} Q_{|\nu\mu|\beta)} \delta g^{\alpha\beta} - 2 Q_{(\alpha}{}^\rho{}_{\beta)} \nabla_\rho \delta g^{\alpha\beta} - 2 \left( Q_\lambda{}^{\mu\nu} + Q^{(\mu\nu)}{}_\lambda \right) \delta \Gamma^\lambda{}_{\nu\mu} \, , \\
		\delta \left( Q^\rho Q_\rho \right) = & \left( Q_\alpha Q_\beta - 2 Q^\rho Q_{\rho\alpha\beta} \right) \delta g^{\alpha\beta} - 2 Q^\rho g_{\alpha\beta} \nabla_\rho \delta g^{\alpha\beta} - 4 Q^{(\rho} \delta^{\mu)}{}_\lambda \delta \Gamma^\lambda{}_{\rho\mu} \, , \\
		\delta \left( Q^\rho \tilde{Q}_\rho \right) = & - \tilde{Q}^\rho Q_{\rho\alpha\beta} \delta g^{\alpha\beta} - \left( \tilde{Q}^\rho g_{\alpha\beta} + Q_{(\alpha} \delta^\rho{}_{\beta)} \right) \nabla_\rho \delta g^{\alpha\beta} - \left[ \left( Q^{(\mu} + \tilde{Q}^{(\mu} \right) \delta^{\rho)}{}_\lambda + Q_\lambda g^{\mu\rho} \right] \delta \Gamma^\lambda{}_{\rho\mu} \, , \\
		\delta \left( \tilde{Q}^\rho \tilde{Q}_\rho \right) = & - \tilde{Q}_\alpha \tilde{Q}_\beta \delta g^{\alpha\beta} - 2 \tilde{Q}_{(\alpha} \delta^\rho{}_{\beta)} \nabla_\rho \delta g^{\alpha\beta} - 2 \left( \tilde{Q}_\lambda g^{\mu\rho} + \tilde{Q}^{(\mu} \delta^{\rho)}{}_\lambda \right) \delta \Gamma^\lambda{}_{\rho\mu} \, .
	\end{align}
\end{subequations}
On the other hand, the vector which gives rise to the standard boundary term in teleparallel theories has the following variational properties
\begin{equation}\label{eq:variation-boundary-vector}
	\delta \left( Q^\mu - \tilde{Q}^\mu \right) = \nabla_\nu \delta g^{\nu\mu} - g_{\alpha\beta} \nabla^\mu \delta g^{\alpha\beta} + \left( Q^\mu{}_{\alpha\beta} + \delta^\mu{}_{(\alpha} Q_{\beta)} \right) \delta g^{\alpha\beta} + 2 \left( \delta^{(\rho}{}_\alpha g^{\mu)\lambda} - g^{\mu\rho} \delta^\lambda{}_\alpha \right) \delta \Gamma^\alpha{}_{\lambda\rho} \, . 
\end{equation}
Furthermore, we define the general non-metricity scalar
\begin{equation}
	\barQ = c_1 Q^{\rho\mu\nu} Q_{\rho\mu\nu} + c_2 Q^{\rho\mu\nu} Q_{\mu\rho\nu} + c_3 Q_\mu Q^\mu + c_4 Q_\mu \tilde{Q}^\mu + c_5 \tilde{Q}_\mu \tilde{Q}^\mu \, ,
\end{equation}
which reduces to the teleparallel equivalent one for $c_1=\tfrac{1}{4}$, $c_2=-\tfrac{1}{2}$, $c_3=-\tfrac{1}{4}$, $c_4 = \tfrac{1}{2}$ and $c_5=0$. The variation of this general scalar expression has the following form
\begin{equation}\label{eq:var-barQ}
	\begin{split}
		\delta \barQ = & \left[ c_1 Q_\alpha{}^{\mu\nu} Q_{\beta\mu\nu} - 2 c_1 Q^{\mu\nu}{}_\alpha Q_{\mu\nu\beta} - 2 c_2 Q^{\mu\nu}{}_\alpha Q_{\nu\mu\beta} + c_3 Q_\alpha Q_\beta - 2 c_3 Q^\rho Q_{\rho\alpha\beta} \right.\\
		& \left. - c_4 \tilde{Q}^\rho Q_{\rho\alpha\beta} - c_5 \tilde{Q}_\alpha \tilde{Q}_\beta \right] \delta g^{\alpha\beta} - \left[ 2 c_1 Q^\rho{}_{\alpha\beta} + 2 c_2 Q_{(\alpha}{}^\rho{}_{\beta)} + \left( 2 c_3 Q^\rho + c_4 \tilde{Q}^\rho \right) g_{\alpha\beta} \right.\\
		& \left. + \left( c_4 Q_{(\alpha} + 2 c_5 \tilde{Q}_{(\alpha} \right) \delta^\rho{}_{\beta)} \right] \nabla_\rho \delta g^{\alpha\beta} - \left[ 2 \left( 2 c_1 + c_2 \right) Q^{(\mu\nu)}{}_\lambda + 2 c_2 Q_\lambda{}^{\mu\nu} \right.\\
		& \left. + \left( 4 c_3 + c_4 \right) Q^{(\mu} \delta^{\nu)}{}_\lambda + \left( c_4 + 2 c_5 \right) \tilde{Q}^{(\mu} \delta^{\nu)}{}_\lambda + c_4 Q_\lambda g^{\mu\nu} \right] \delta \Gamma^\lambda{}_{\nu\mu} \, .
	\end{split}
\end{equation}
Whence, the metric field equations are found to be
\begin{equation}
	\begin{split}
		& c_1 Q_\alpha{}^{\mu\nu} Q_{\beta\mu\nu} - 2 c_1 Q^{\mu\nu}{}_\alpha Q_{\mu\nu\beta} - 2 c_2 Q^{\mu\nu}{}_\alpha Q_{\nu\mu\beta} + \left( \left( c_1 -2c_3 \right) Q^\rho - c_4 \tilde{Q}^\rho \right) Q_{\rho\alpha\beta} \\
		& + c_2 Q_\rho Q_{(\alpha}{}^\rho{}_{\beta)} + \left( c_3 + \frac{c_4}{2} \right) Q_\alpha Q_\beta - c_5 \tilde{Q}_\alpha \tilde{Q}_\beta + c_5 Q_{(\alpha} \tilde{Q}_{\beta)} \\
		& - \frac{1}{2} g_{\alpha\beta} \left( c_1 Q^{\rho\mu\nu} Q_{\rho\mu\nu} + c_2 Q^{\rho\mu\nu} Q_{\mu\rho\nu} - c_3 Q_\rho Q^\rho + c_5 \tilde{Q}_\rho \tilde{Q}^\rho  \right) \\
		& + \nabla_\rho \left( 2 c_1 Q^\rho{}_{\alpha\beta} + 2 c_2 Q_{(\alpha}{}^\rho{}_{\beta)} + \left( 2 c_3 Q^\rho + c_4 \tilde{Q}^\rho \right) g_{\alpha\beta} + \left( c_4 Q_{(\alpha} + 2 c_5 \tilde{Q}_{(\alpha} \right) \delta^\rho{}_{\beta)} \right) = 0 \, .
	\end{split}
\end{equation}

Now we focus on the $3+1$ decomposition of the same scalar expression that we have just considered. By repeatedly using the completeness relations Eqs.\ \eqref{eq:completeness-non-metricity}, \eqref{eq:completeness-first-trace} and \eqref{eq:completeness-second-trace} we find
\begin{equation}
	\begin{split}
		\mathbb{\overline{Q}} = & (2 c_1 -  c_2) \epsilon A_{ab} A^{ab} + 4 (c_1 + c_2 + c_3 + c_4 + c_5) \epsilon \alpha^2 + \mathbb{\overline{Q}}{}^{(3)} + 2 c_5 \epsilon \tilde{Q}^{(3)}{}_a \lambda^{a} \\
		& + (2 c_1 + c_2 + c_5) \lambda_{a} \lambda^{a} + (2 c_2 + c_4) \lambda^{a} \theta_{a} + \epsilon \left( 2 c_3 Q^{(3)}{}_a + c_4 \tilde{Q}^{(3)}{} \right) \theta^a + (c_1 + c_3) \theta_{a} \theta^{a} \\
		& + 2 (c_4 + 2 c_5) \alpha S^{a}{}_{a} + (2 c_1 + c_2) \epsilon S_{ab} S^{ab} + c_5 \epsilon S^{a}{}_{a} S^{b}{}_{b} + 2 c_2 \epsilon S^{ab} \Sigma_{ab} + 2 (2 c_3 + c_4) \alpha \Sigma^{a}{}_{a} \\
		& + c_1 \epsilon \Sigma_{ab} \Sigma^{ab} + c_4 \epsilon S^{a}{}_{a} \Sigma^{b}{}_{b} + c_3 \epsilon \Sigma^{a}{}_{a} \Sigma^{b}{}_{b} + 2 (2 c_1 + c_2 + c_5) \lambda^{a} \kappa_{a} + (2 c_2 + c_4) \theta^{a} \kappa_{a} \\
		& + \epsilon \left( c_4 Q^{(3)}{}_{a} + 2 c_5 \tilde{Q}^{(3)}{}_{a} \right) \kappa^{a} + (2 c_1 + c_2 + c_5) \kappa_{a} \kappa^{a} \, .
	\end{split}
\end{equation}
In the teleparallel limit this expression becomes much simpler:
\begin{equation}
	\begin{split}
		\mathbb{\overline{Q}} \overset{telep}{=} & 4 (c_1 + c_2 + c_3 + c_4 + c_5) \epsilon \alpha^2 + \mathbb{\overline{Q}}{}^{(3)} + \epsilon \left( 2 c_3 Q^{(3)}{}_a + c_4 \tilde{Q}^{(3)}{} \right) \theta^a + (c_1 + c_3) \theta_{a} \theta^{a} \\
		& + 2 (2 c_3 + c_4) \alpha \Sigma^{a}{}_{a} + c_1 \epsilon \Sigma_{ab} \Sigma^{ab} + c_3 \epsilon \Sigma^{a}{}_{a} \Sigma^{b}{}_{b} \, .
	\end{split}
\end{equation}
Notice that in the equivalent case the extrinsic non-metricity scalar $\alpha$ identically decouples from the theory, thus yielding the premises for the formal proof of the equivalence with GR.


\bibliographystyle{chetref}


\end{document}